\pdfoutput=1
\documentclass[a4paper,11pt]{article}

\usepackage{jcappub}

\usepackage[cp1250]{inputenc}
\usepackage{graphicx}
\usepackage{epstopdf}
\usepackage{amsmath,amsthm,latexsym,amssymb,amsfonts,epsfig}
\usepackage{fontenc,layout}
\usepackage{hyperref}
\usepackage[usenames,dvipsnames]{xcolor}

\newcommand{\be}{\begin{equation}}
\newcommand{\ee}{\end{equation}}
\newcommand{\bea}{\begin{eqnarray}}
\newcommand{\eea}{\end{eqnarray}}
\newcommand{\beq}{\begin{eqnarray}}
\newcommand{\eeq}{\end{eqnarray}}

\numberwithin{equation}{section}

\title{Gravitational wave energy budget in strongly supercooled phase transitions}

\author[1,2,3]{John~Ellis,}
\author[1]{Marek~Lewicki,}
\author[4]{Jos\'e~Miguel~No,}
\author[1]{Ville~Vaskonen}
\affiliation[1]{Department of Physics, King's College London, Strand, WC2R 2LS London, UK}
\affiliation[2]{Theoretical Physics Department, CERN, Geneva, Switzerland}
\affiliation[3]{National Institute of Chemical Physics \& Biophysics, R\"avala 10, 10143 Tallinn, Estonia}
\affiliation[4]{Departamento de F\'isica Te\'orica and Instituto de F\'isica Te\'orica, IFT-UAM/CSIC,
Cantoblanco, 28049, Madrid, Spain}
\emailAdd{john.ellis@cern.ch}
\emailAdd{marek.lewicki@kcl.ac.uk}
\emailAdd{josemiguel.no@uam.es}
\emailAdd{ville.vaskonen@kcl.ac.uk}

\abstract{\\
We derive efficiency factors for the production of gravitational waves through bubble collisions and plasma-related sources in strong phase transitions, and find the conditions under which the bubble collisions can contribute significantly to the signal. We use lattice simulations to clarify the dependence of the colliding bubbles on their initial state. We illustrate our findings in two examples, the Standard Model with an extra $|H|^6$ interaction and a classically scale-invariant $U(1)_{\rm B-L}$ extension of the Standard Model. The contribution to the GW spectrum from bubble collisions is found to be negligible in the $|H|^6$ model, whereas it can play an important role in parts of the parameter space in the scale-invariant $U(1)_{\rm B-L}$ model. In both cases the sound-wave period is much shorter than a Hubble time, suggesting a significant amplification of the turbulence-sourced signal. We find, however, that the peak of the plasma-sourced spectrum is still produced by sound waves with the slower-falling turbulence contribution becoming important off-peak.
\\
\\
\\
\\
\\
KCL-PH-TH/2019-32, CERN-TH/2019-032, IFT-UAM/CSIC-19-32}

\begin{document}

\maketitle

\section{Introduction}

A first-order phase transition is a common feature of particle physics models. Though absent in the Standard Model (SM), such a transition would occur in many proposed extensions. Interest in this possibility was motivated traditionally by the hope of realizing electroweak baryogenesis~\cite{Kuzmin:1985mm,Cohen:1993nk,Riotto:1999yt,Morrissey:2012db}. However, more recently such scenarios have enjoyed renewed attention due to the observational prospects associated with a gravitational wave (GW) background that such a transition could produce~\cite{Kamionkowski:1993fg,Apreda:2001us,Grojean:2006bp,Huber:2008hg,Espinosa:2008kw,Dorsch:2014qpa,Kakizaki:2015wua,Jaeckel:2016jlh,
Dev:2016feu,Hashino:2016rvx,Chala:2016ykx,Chala:2018opy,Artymowski:2016tme,Hashino:2016xoj,
Vaskonen:2016yiu,Dorsch:2016nrg,Beniwal:2017eik,
Marzola:2017jzl,Kang:2017mkl,Chala:2018ari,Huang:2016odd,Huang:2017kzu,Hashino:2018zsi,Vieu:2018zze,Huang:2018aja,Bruggisser:2018mrt,Megias:2018sxv,Croon:2018erz,Alves:2018jsw,Baratella:2018pxi,Angelescu:2018dkk,Croon:2018kqn,Miura:2018dsy,Brdar:2018num,Mazumdar:2018dfl,Beniwal:2018hyi,Marzo:2018nov,Baldes:2018emh,Prokopec:2018tnq,Fairbairn:2019xog}~\footnote{Another interesting possibility is the generation of magnetic fields during such a transition, see, e.g.,~\cite{Zhang:2019vsb} or production of primordial black holes~\cite{Hawking:1982ga,Moss:1994iq,Konoplich:1999qq}.}.

Several years ago leading-order calculations encouraged optimism about a possible GW signal from a first-order transition due to collisions between runaway bubble walls~\cite{Bodeker:2009qy} that keep accelerating throughout their existence as the vacuum pressure driving the expansion could overcome the friction enacted by the plasma, which was constant in a leading-order calculation. However, more recently a next-to-leading order calculation~\cite{Bodeker:2017cim} has shown that at this order the friction is proportional to the gamma factor of the wall, and hence the bubble walls cannot run away. However, significant dilution of the surrounding plasma could still allow the walls to accelerate for a prolonged period of time and carry a significant amount of energy. 

In this work we quantify conditions necessary for generating observable GW signals from bubble-wall collisions in first-order phase transitions. We also calculate the energy budget of such strong transitions and provide the efficiency factors that allow one to predict accurately the relative strengths of the sources associated with the plasma and the bubble walls themselves. Moreover, we re-examine the efficiencies of the plasma sources, taking account of the fact that the sound-wave period was found in~\cite{Ellis:2018mja} to be shorter than a Hubble time, leading to a reduction of the sound-wave contribution as compared to previous estimates. At the same time, under the assumption that the available plasma sound wave kinetic energy is entirely converted into turbulent motion once the sound wave period ends, the GW signal from turbulence is greatly amplified and could be comparable to that from sound waves. However, we still find that the peak of the spectrum in scenarios where the plasma sources dominate the signal is produced by sound waves, with the slower-falling turbulence contribution becoming more important off-peak.

We then discuss in detail two examples that illustrate our results for polynomial and conformal potentials. The polynomial example is an extension of the SM scalar potential parametrised by a non-renormalisable six-dimensional Higgs field operator. In this case, we confirm that significant supercooling is not possible, in agreement with~\cite{Ellis:2018mja}, and thus the bubble collision signal is negligible. We then further quantify the interplay of sound waves and turbulence in this model.
The second example corresponds to a conformal scenario, namely an extension of the SM by a spontaneously broken $U(1)_{\rm B-L}$ gauge symmetry featuring scale invariance at the classical level, in which strong supercooling can be realised. We find the amount of supercooling required to produce a significant GW signal from bubble collisions, and discuss the interplay between bubble collisions and plasma GW sources in this case.

We also re-evaluate the equation of state during the transition. We take into account the fraction of the vacuum energy converted into bubble wall kinetic energy, and account for its redshifting as radiation to compute more accurately the expansion rate. We find that this modification does not influence significantly the transition up to the percolation time.
The other significant modification of the equation of state occurs in the conformal model, in which the very small coupling to the plasma can cause the field to oscillate around the minimum of the potential after the transition for an extended period of time before it reheats the Universe. This leads to a short period of evolution resembling matter domination in parts of the parameter space.  

The paper is organised as follows. In Section~\ref{sec:wallenergy} we start by calculating the bubble wall energy in the thin-wall approximation, and then perform lattice simulations to extend the thin-wall result to realistic initial bubble profiles. We give expressions for the efficiency factors for GW production in Section~\ref{sec:gws}, and study the effect of the bubble wall energy on the expansion rate of the universe in Section~\ref{sec:expansion}. Finally, we apply our results in the two above mentioned model examples in Section~\ref{sec:examples}, and summarize our key conclusions in Section~\ref{sec:conclusions}.

\section{Energy stored in bubble walls}
\label{sec:wallenergy}

\subsection{Thin-wall bubbles}

 We start with the thin-wall approximation~\cite{Coleman:1977py}, in which the bubble wall is treated as a boundary of negligible width between the two phases. With this assumption one can write down a simple Lagrangian as a function of the bubble size $R$ {(see, e.g.,~\cite{Darme:2017wvu})}:
\be \label{eq:lagr}
\mathcal{L}= -4\pi \sigma R^2 \sqrt{1-\dot{R}^2}  + \frac{4\pi}{3} R^3 p \, ,
\ee
where $\sigma$ is the bubble wall tension and $p$ is the pressure acting on the bubble wall. {In the case of a vacuum transition} the pressure would simply come from the vacuum energy difference $p = \Delta V = V_f - V_t$. The corresponding total energy derived is given by 
\be \label{eq:bubble_tw_energy}
\mathcal{E} = 4\pi \gamma \sigma R^2 - \frac{4\pi}{3} R^3 p\,,
\ee
where $\gamma = 1/\sqrt{1-\dot{R}^2}$ is the Lorentz gamma factor of the bubble wall. Typically, in order to calculate the temperature at which bubbles that can drive the transition begin to nucleate, one finds the action of the smallest bubble that will not collapse. This {means} looking for a static solution by setting $\gamma=1$ and finding the {critical} radius for which ${\rm d}\mathcal{E}/{\rm d}R=0$, which gives~\footnote{This is the usual result for a $O(3)$ symmetric bubble in thin-wall approximation~\cite{Linde:1981zj}.} $R = 2\sigma/p \equiv R_c$. 
Tracking the bubble growth after nucleation {then} requires using the full equation of motion from the Lagrangian~\eqref{eq:lagr},
\be \label{eq:bubble_eom}
\ddot{R}+2\frac{1-\dot{R}^2}{R}=\frac{p}{\sigma}\left( 1-\dot{R}^2 \right)^\frac{3}{2} \,,
\ee
which indeed has a static solution for $R=R_c$. For an expanding bubble, the initial radius of the bubble $R_0$ has to be larger than this critical radius.

It is convenient to rewrite the equation of motion~\eqref{eq:bubble_eom} in terms of the $\gamma$ factor:
\be
\frac{{\rm d}\gamma}{{\rm d}R} + \frac{2\gamma}{R} = \frac{p}{\sigma} \,.
\ee
This can easily be solved analytically, with the initial condition $\gamma(R_0)=1$ yielding
\be \label{eq:gammasol}
\gamma = \frac{pR}{3\sigma} + \frac{R_0^2}{R^2} - \frac{p R_0^3}{3\sigma R^2} \approx \frac{2R}{3R_0} + \frac{R_0^2}{3R^2} \,,
\ee
where in the last step we {have} assumed that the initial radius is only slightly larger than the critical one, $R_0\approx R_c$. Using the above equation in the bubble energy~\eqref{eq:bubble_tw_energy}, we can clearly see that the second term simply amounts to adding a constant to the energy, which corresponds to the initial action of the nucleated bubble. 
{We see from Eq.~\eqref{eq:gammasol} that} the bubble acceleration will simply follow $\gamma \approx 2R/(3R_0)$ as the bubble grows~\footnote{{We will verify this approximation in Section~\ref{sec:lattice} via a comparison with the numerical evolution of realistic bubble profiles obtained by solving numerically the full equations of motion of the fields in a realistic potential.}}.   

Whilst inclusion of the friction force exerted by the plasma surrounding the bubble is highly non-trivial in general~\cite{Dorsch:2018pat}, the case of {very relativistic bubble walls (as expected in strongly supercooled phase transitions)} can be tackled using the approximation above {with the addition of} two extra terms in the friction. The leading-order term, at a given temperature, is just a constant~\cite{Bodeker:2009qy}
\be 
\label{eq:frictionLO}
\Delta P _{\rm LO} \approx \frac{\Delta m^2 T^2}{24}\,,
\ee
and the next-to-leading order term associated with particle splitting/transition radiation at the bubble  wall~\cite{Bodeker:2017cim} is proportional to the Lorentz $\gamma$ factor of the wall:
\be 
\label{eq:frictionNLO}
\gamma \Delta P _{\rm NLO} \approx \gamma \,g^2 \Delta m_V T^3 \,.
\ee
 The differences in (squared) masses between the symmetric and the broken phases are given by
\be \label{eq:Deltam}
\Delta m^2 \equiv \sum_i c_i N_i \Delta m_i^2 \,, \quad 
g^2\Delta m_V \equiv \sum_{i\in V} g_i^2 N_i \Delta m_i
\ee
where $N_i$ is the number of internal degrees of freedom of particle $i$, $c_i=1 \, (1/2)$ for bosons (fermions), $\Delta m_i^2 = m_{i,t}^2-m_{i,f}^2$\,, $\Delta m_i = m_{i,t}-m_{i,f}^2$\,, the sums run over particles that gain mass in the transition, and in the latter case only gauge bosons are included, where the $g_i$ are their respective gauge couplings.

\begin{figure}
\begin{center}
\includegraphics[width=0.75\textwidth]{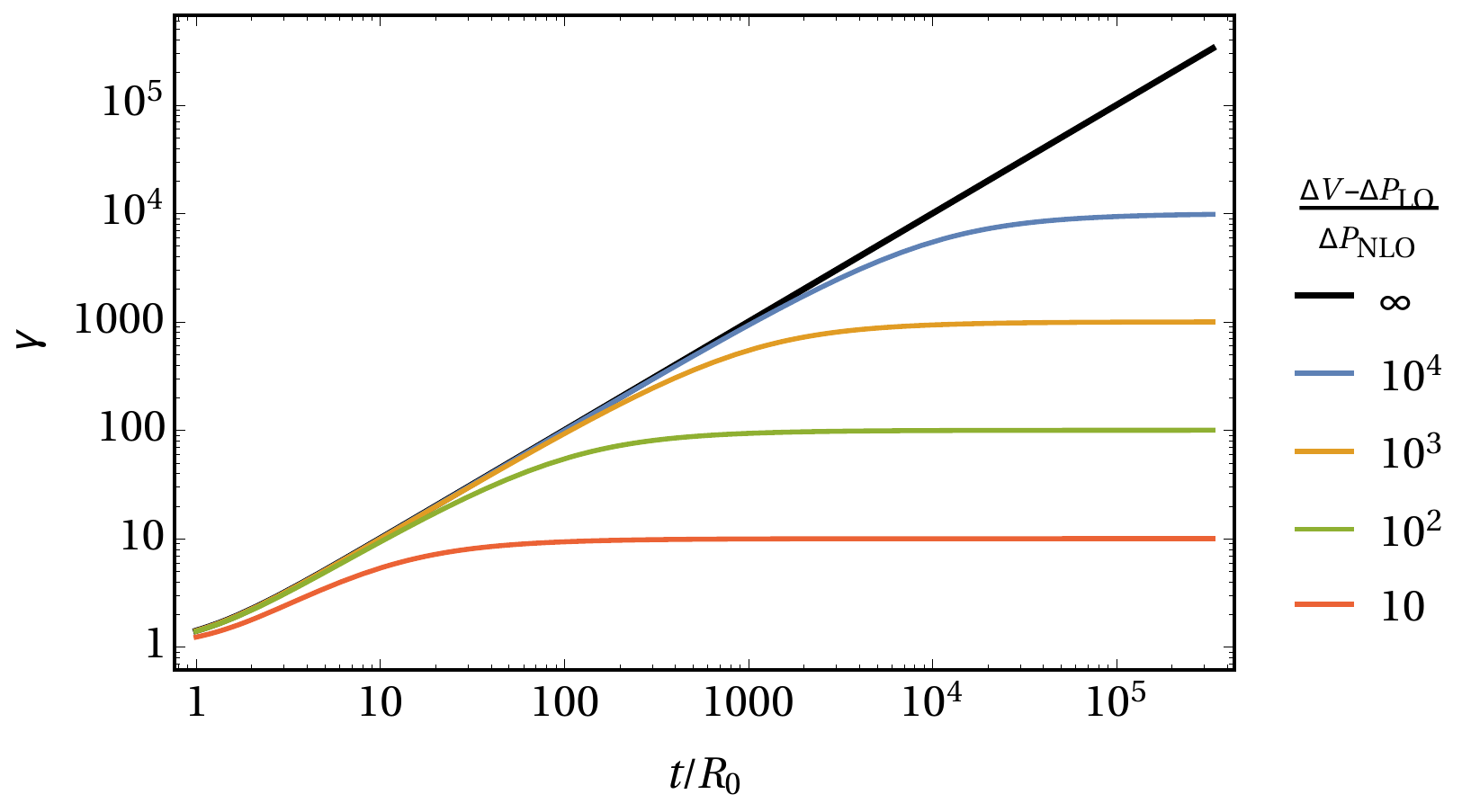}
\caption{ The Lorentz $\gamma$ factor of the wall of an expanding bubble as a function of time. Different lines correspond to different values of $(\Delta V - \Delta P_{\rm LO})/\Delta P_{\rm NLO}$.}
\label{fig:gammagrowth}
\end{center}
\end{figure}

We can include the {friction terms \eqref{eq:frictionLO} and \eqref{eq:frictionNLO}} in the equation of motion~\eqref{eq:bubble_eom} by writing
\be \label{eq:pressure}
p \equiv \Delta V - \Delta P _{\rm LO} - \gamma \Delta P _{\rm NLO}\,.
\ee
We {note that the above simplified treatment holds provided that $\Delta V>\Delta P_{\rm LO}$ (known previously as {\sl runaway} 
behaviour~\cite{Bodeker:2009qy})}~\footnote{{Otherwise it would yield nonsensical solutions, i.e.,~bubbles contracting under the plasma pressure, which would not persist in a more complete thermodynamical treatment.}}. We can now solve numerically the equation of motion~\eqref{eq:bubble_eom} with the pressure term~\eqref{eq:pressure} to find the evolution of $\gamma$ with time, as shown in Fig.~\ref{fig:gammagrowth}.

The results shown in Fig.~\ref{fig:gammagrowth} depend only on the ratio of $\Delta V - \Delta P _{\rm LO}$ to $\Delta P _{\rm NLO}$. As expected, $\gamma$ {ceases} to grow when the ratio ($\Delta V - \Delta P _{\rm LO})/(\gamma \Delta P _{\rm NLO})$ is equal to one. Using the analytical solution~\eqref{eq:gammasol} for $R\gg R_0$, truncated when the friction terms equilibrate, i.e. for
\be \label{eq:gammaeq}
\gamma_{\rm eq} \equiv \frac{\Delta V - \Delta P _{\rm LO}}{\Delta P _{\rm NLO}} \,,
\ee
we find that the radius at which bubble walls cease to accelerate is $R_{\rm eq} \equiv 3\gamma_{\rm eq} R_0/2$. 

\vspace{1mm}

{The} GW signal from bubble collisions may be significant if the bubble size at percolation $R_*$ is not much bigger than $R_{\rm eq}$, {as} this would mean that a significant {fraction of the} vacuum energy was used to accelerate the bubble walls. {We can easily approximate that energy fraction as follows:} The bubble expands {at} roughly the speed of light until it reaches its final size at collision $R=R_*$. Up until the bubble reaches $R_{\rm eq}$, {essentially} all the energy from vacuum conversion is used to accelerate the wall, except for the part used to overcome the leading-order friction, i.e., $\Delta V -\Delta P_{\rm LO}$. The amount of energy gained from the vacuum conversion is simply $\Delta V$ times the volume, thus the fraction of the total energy that is stored in the wall is {given by}
\be \label{eq:Ewall}
\frac{E_{\rm wall}}{E_{V}}=\frac{\Delta V-\Delta P_{\rm LO}}{\Delta V} \,.
\ee
After the bubble reaches $R_{\rm eq}$ it {stops accelerating}, and only grows at a constant rate. In that case the fraction of the total energy released that is stored in the wall is
\be 
\frac{E_{\rm wall}}{E_{V}} = \frac{(\Delta V-\Delta P_{\rm LO}) \frac{4\pi}{3} R_{\rm eq}^3}{\Delta V \frac{4\pi}{3} R_{\rm *}^3} + \frac{4\pi \gamma_{\rm eq}\sigma (R_*^2-R_{\rm eq}^2)}{\Delta V \frac{4\pi}{3} R_{\rm *}^3} \,,
\ee
where the first term on the {right-hand} side accounts for all energy gained before the terminal velocity is reached, and the second term takes into account the increase in the wall area when the bubble expands at constant velocity. Using $3\gamma_{\rm eq}\sigma = R_{\rm eq} \Delta V$ we get
\be \label{eq:Ewall}
\begin{split}
\frac{E_{\rm wall}}{E_{V}}
&= \frac32 \gamma_{\rm eq} \frac{R_0}{R_*}\left[1-\frac{\Delta P_{\rm LO}}{\Delta V}\left(\frac32\gamma_{\rm eq}\frac{R_0}{R_*}\right)^2\right]\, .
\end{split}
\ee
The standard treatment in the literature is to define
\be \label{eq:alpha}
\alpha \equiv \frac{1}{\rho_R} \left(\Delta V - \frac{T}{4} \Delta \frac{{\rm d}V}{{\rm d} T} \right) \simeq \frac{\Delta V}{\rho_R}\,,
\ee
where $\rho_R$ is the radiation energy density and the last approximation holds in the supercooled case that is of {interest} to us. We also {define~\cite{Espinosa:2010hh} (see also~\cite{Caprini:2015zlo})}
\be \label{eq:alpha_infty}
\alpha_\infty \equiv \frac{\Delta P_{\rm LO}}{\rho_R} =  \frac{1}{24}\frac{\Delta m^2 T^2}{\rho_R} \,,
\ee
which via $\alpha = \alpha_\infty$ determines the weakest transition for which the vacuum conversion pressure driving the bubble expansion is larger than the leading-order plasma {friction~\eqref{eq:frictionLO}. As discussed above, the validity of our treatment requires $\alpha > \alpha_\infty$, as for weaker cases} one should instead try to find the wall velocity in the plasma background, {$v_{w}$}, which can be much smaller than the speed of light~\cite{Dorsch:2018pat}. 
{Analogously to~\eqref{eq:alpha_infty}}, we can define
\be \label{eq:alpha_prime_infty}
\alpha_{\rm eq} \equiv \frac{\Delta P_{\rm NLO}}{\rho_R} = \frac{g^2 \Delta m_V T^3}{\rho_R} \,,
\ee
which allows us to rewrite the Lorentz $\gamma$ factor for the terminal velocity of the wall~\eqref{eq:gammaeq} as
\be \label{eq:gammaeq2}
\gamma_{\rm eq} = \frac{\alpha-\alpha_\infty}{\alpha_{\rm eq}} \,.
\ee
Finally, by defining $\gamma_*$ as the gamma factor that the bubble wall would reach if the {next-to-leading order} friction term $\gamma \Delta P_{\rm NLO}$ were neglected,
\be \label{eq:gamma_star}
\gamma_* \equiv \frac{2}{3} \frac{R_*}{R_0}\,,
\ee
the fraction of the total energy at percolation that is stored in the wall can be written as
\be \label{eq:Ewall}
\frac{E_{\rm wall}}{E_V} = 
\begin{cases}
\frac{\gamma_{\rm eq}}{\gamma_*} \left[1-\frac{\alpha_\infty}{\alpha}\left(\frac{\gamma_{\rm eq}}{\gamma_*}\right)^2 \right] \,, \quad &\gamma_*>\gamma_{\rm eq} \\
1-\frac{\alpha_\infty}{\alpha} \,, \quad &\gamma_*\leq \gamma_{\rm eq} \,.
\end{cases}
\ee
As expected, the fraction of the total energy that goes into the bubble wall {quickly decreases once the wall reaches its terminal velocity}. The rest of the vacuum energy, $E_V - E_{\rm wall}$, goes into {kinetic and thermal energy of the plasma around} the bubble wall. We will explore the implications of these results for the resulting GW spectra in {Section~\ref{sec:gws}.}

\subsection{Numerical simulation}
\label{sec:lattice}

Generically, the initial bubble wall after nucleation does not resemble that obtained from the thin-wall approximation. However, the wall gets thinner as the bubble expands and accelerates, and eventually the thin-wall analysis described above {does} hold. In order to study the {bubble} evolution at early times we resort to a numerical simulation of an expanding spherically symmetric bubble. 
We simulate one bubble starting from a initial profile $\phi_{\rm in}(r)$ obtained, as usual, by minimizing the $O(3)$ symmetric action
\be \label{eq:S3}
S_3 = 4\pi \int r^2 {\rm d}r \left[\frac{1}{2}\left(\frac{{\rm d}\phi}{{\rm d}r}\right)^2 + V(\phi) \right] \,.
\ee
The initial profile $\phi_{\rm in}(r)$ has to satisfy the conditions ${\rm d}\phi_{\rm in}/{\rm d}r = 0$ at $r=0$ and $\phi_{\rm in}\to 0$ at $r\to \infty$. The evolution of a three-dimensional spherically-symmetric bubble can be described using only the time and radial coordinate:
\be
-\partial_t^2 \phi + \partial_r^2 \phi + \frac{2}{r}\partial_r \phi = \frac{{\rm d}V(\phi)}{{\rm d}\phi}\, 
\ee
with the initial conditions $\phi(t=0,r) = \phi_{\rm in}(r)$ and $\partial_t \phi(t,r) = 0$. For the scalar potential {$V(\phi)$}, we use that discussed later in Section~\ref{sec:csi}. However, we note that the qualitative results do not depend on the actual form of the potential.

\begin{figure}
\begin{center}
\includegraphics[width=0.75\textwidth]{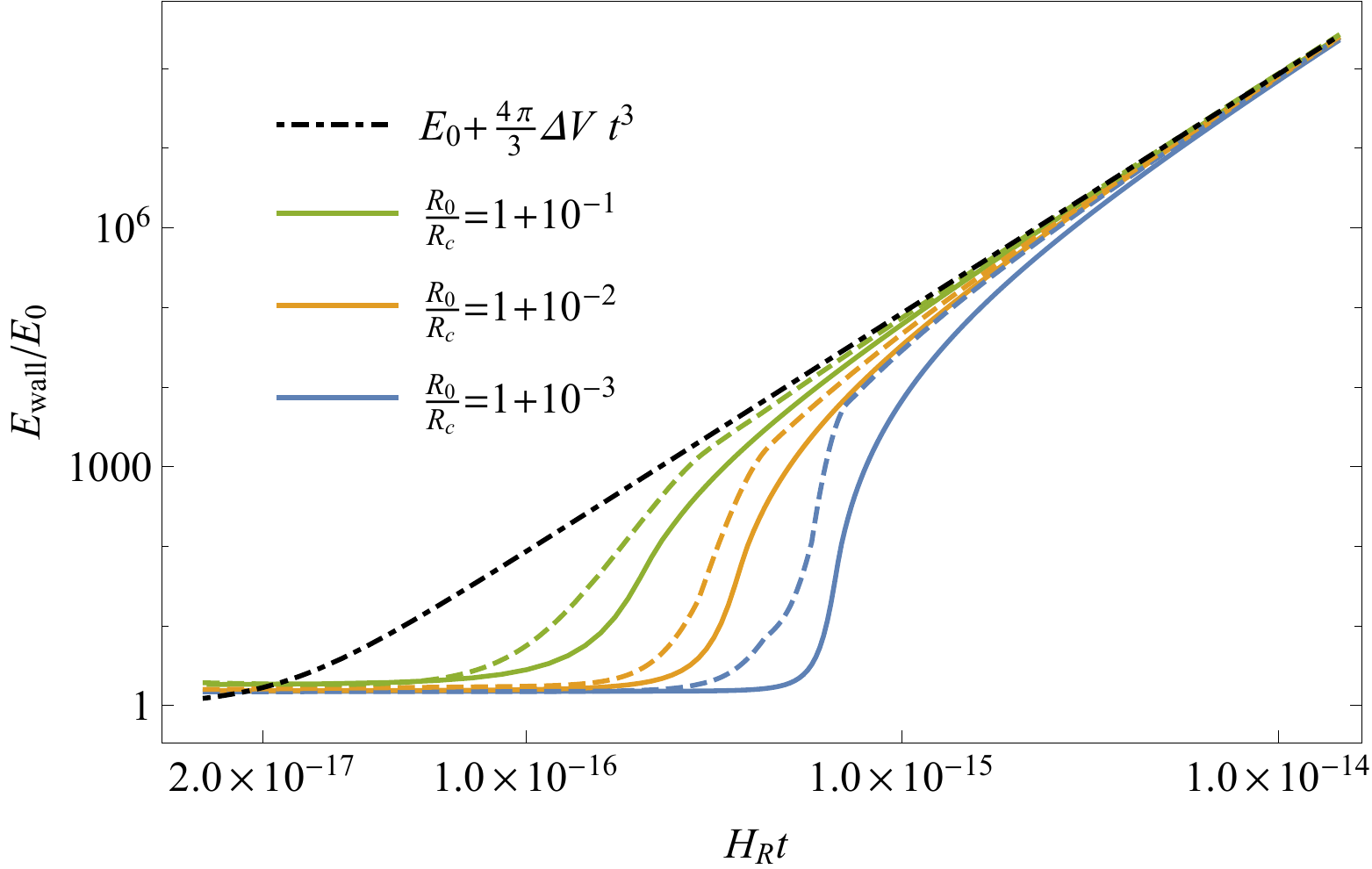}
\caption{ The solid lines show the energy stored in the bubble wall {$E_{\rm wall}$} as a function of time calculated through a lattice evolution for three different initial energies. The dashed lines show the corresponding results obtained by the approximation~\eqref{eq:Ewall2} with $R(t)$ solved from~\eqref{eq:gammasol}, and the dot-dashed one shows the simplest possible approximation~\eqref{eq:Ewall2} with $R(t)=t$ and $\Delta V(t) = \Delta V$.}
\label{fig:Ewallplot}
\end{center}
\end{figure}

{Using} energy conservation, we can describe the energy stored in the wall through the potential energy of the bubble, thus avoiding the use of the {wall tension in the thin-wall limit} $\sigma$ (and its inherent inaccuracy). We find that {the energy stored in the bubble wall is well approximated by}
\be \label{eq:Ewall2}
E_{\rm wall} = E_0 + \frac{4\pi}{3} R(t)^3 \Delta V(t) \,,
\ee
where $E_0 = S_3(\phi = \phi_{\rm in})$ is the initial total energy of the bubble and $\Delta V(t) = V(0) - V(\phi(t,r=0))$ is the potential energy difference between the center of the bubble $r=0$ and the outside $r\to\infty$. This accounts for the fact that the field does not reach the minimum of the potential inside the initial bubble. The radius $R(t)$ can be solved numerically from Eq.~\eqref{eq:gammasol}, with $R_0$ being the radius of a thin-wall bubble with the potential energy corresponding to the one of the actual initial bubble,
\be \label{eq:R0}
R_0 \equiv \left[\frac{3 E_{0,V}}{4\pi\Delta V(t=0)}\right]^{1/3} \,,
\ee
where $E_{0,V}$ is the potential energy contribution to {the initial bubble energy $E_0$.} 

We compare the results from our {numerical} study with those using the  {approximation~\eqref{eq:Ewall2}} in Fig.~\ref{fig:Ewallplot}. While bubbles with energy infinitesimally larger than the critical value can in principle take a very long time before they start growing, the bubble wall energy always reaches asymptotically the dot-dashed line corresponding to~\eqref{eq:Ewall2} with $R(t)=t$ and $\Delta V(t) = \Delta V$ that provides the simplest possible approximation of the {bubble} wall energy. Therefore, the thin-wall prescription presented in the earlier part of this Section {is justified and the} initial bubble radius should be calculated using Eq.~\eqref{eq:R0}.

\section{Gravitational wave signals}
\label{sec:gws}

{{We now give expressions for the efficiency factors for the various phase transition sources to generate GWs, as well as} the resulting GW spectra. {We} focus on {strong} transitions with {very} relativistic {bubble} wall velocities {$v_{w} \to 1$, but note that} the expressions {given below for the GW spectra are more general}~\footnote{{Our results can be generalized to smaller wall velocities by neglecting the bubble collision efficiency factor $\kappa_{\rm col}$ from~\eqref{eq:kappa_col} and using the corresponding plasma efficiency factor $\kappa_{\rm sw}$ in the case of smaller wall velocities as given in~\cite{Espinosa:2010hh}.}}.}
{These expressions} are given at the time of percolation and we describe the factors necessary to obtain the GW spectrum that would be observed today in {Section~\ref{sec:expansion}, via} Eqs.~\eqref{eq:Omegaredshift} and~\eqref{eq:fredshift}.

\vspace{1mm}

As shown in Section~\ref{sec:wallenergy}, the fraction $E_{\rm wall}/E_V$ of the vacuum energy goes into accelerating the bubble wall
{(note,} however, that, as shown in Section~\ref{sec:lattice}, the thin-wall value for $R_0$ does not fit well the bubble wall energy, but one should rather use $R_0$ defined via {Eq.~\eqref{eq:R0}), which we define as the efficiency factor for bubble collisions as a source of GW:}
\be
\label{eq:kappa_col}
\kappa_{\rm col} \equiv \frac{E_{\rm wall}}{E_V} \,.
\ee
The contribution to the GW signal from the bubble collisions is then given by~\cite{Cutting:2018tjt}
\be
\Omega_{{\rm col},*} = 
0.024\,(H_* R_*)^2 \left(\frac{\kappa_{\rm col}\alpha}{1+\alpha}\right)^2 
\left(\frac{f_*}{f_{\rm col}}\right)^3\left[1+2\left(\frac{f_*}{f_{\rm col}}\right)^{2.07}\right]^{-2.18} \,,
\ee
where\footnote{The bubble collision signal can only be sizable if the walls overcome the leading order plasma friction. This implies $v_w\approx1$ which we already used in the GW signals from bubble collisions here.}
\be
f_{\rm col} = 0.51 R_*^{-1}  
\ee
is the peak frequency of the spectrum, $T_{\rm reh}$ is the plasma temperature after the vacuum energy has decayed (see Section~\ref{sec:expansion}), and $H_*$ is the Hubble rate at percolation~\footnote{Notice that the peak frequency today, given after redshifting by Eq.~\eqref{eq:fredshift}, depends on $R_*$ only through the product $H_* R_*$.}.

The remaining part, $1-E_{\rm wall}/E_{V}$, of the released vacuum energy goes into the surrounding plasma. This gives rise to sound waves propagating in the plasma after the transition that source GWs~\cite{Hindmarsh:2013xza,Hindmarsh:2015qta,Hindmarsh:2016lnk,Hindmarsh:2017gnf}. {As discussed in Section~\ref{sec:wallenergy}, we focus here on the regime $\alpha > \alpha_\infty$,} in which the vacuum pressure accelerating the bubble {wall} overcomes the leading-order friction term {from Eq.~\eqref{eq:frictionLO}, and $v_w \approx 1$.} The efficiency coefficient for sound wave GW production {is then approximately given by~\cite{Espinosa:2010hh,Caprini:2015zlo}}
\be
\kappa_{\rm sw} =\frac{\alpha_{\rm eff}}{\alpha} \frac{\alpha_{\rm eff}}{0.73+0.083\sqrt{\alpha_{\rm eff}}+\alpha_{\rm eff}} \quad, \quad {\rm with} \,\,\,\,\, \alpha_{\rm eff} = \alpha(1-\kappa_{\rm col}) \,.
\ee
The sound wave GW spectrum may then be expressed as~\cite{Caprini:2015zlo} (see also~\cite{Ellis:2018mja})
\be \label{eq:Omegasw}
\Omega_{{\rm sw},*} = 
0.38 
(H_*R_*) (H_*\tau_{\rm sw}) \left(\frac{\kappa_{\rm sw} \,\alpha }{1+\alpha }\right)^2
\left(\frac{f_*}{f_{\rm sw}}\right)^3 \left[1+\frac{3}{4} \left(\frac{f_*}{f_{\rm sw}}\right)^2\right]^{-\frac72} \,,
\ee
and peaks at the frequency~\footnote{The factor $(v_w-c_s)$ in~\eqref{eq:frequencySW} reflects the fact that the characteristic length scale for sound waves is the size of the plasma shell rather than the bubble size~\cite{Hindmarsh:2016lnk}.}
\be
f_{\rm sw} \,=\,  3.4((v_w-c_s)  \, R_*)^{-1} 
\, ,
\label{eq:frequencySW}
\ee
with the speed of sound in the plasma $c_s=1/\sqrt{3}$. In our relativistic wall case $v_w\approx 1$ such that $ (v_w-c_s)\approx 0.422 $. Our results coincide with the standard {GW} prescription for the original {\sl runaway} scenario~\footnote{Prior to the computation of the next-to-leading order friction term~\cite{Bodeker:2017cim}.} (see~\cite{Caprini:2015zlo}) if the bubble walls keep accelerating up to their collision. At the same time, we take into account the energy fraction deposited in plasma if the walls reach a terminal velocity prior to bubble collisions. The term $H_* \tau_{\rm sw}$ in~\eqref{eq:Omegasw}, where $\tau_{\rm sw}$ is the length of the sound wave period,
\be
\tau_{\rm sw} \equiv \min\left[\frac{1}{H_*},\frac{R_*}{U_f}\right] \,.
\ee
accounts for the fact that if sound waves as a GW source are active for less than a Hubble time, the GW amplitude from sound waves appropriately is reduced~\cite{Ellis:2018mja}. The root-mean-square (RMS) fluid velocity $U_f$ can be approximated as~\cite{Hindmarsh:2015qta}
\be
U_f^2 \simeq \frac{3}{v_w(1+\alpha)} \int_{c_s}^{v_w} {\rm d}\xi\, \frac{\xi^2 v(\xi)^2}{1-v(\xi)^2} \simeq \frac34 \frac{\alpha_{\rm eff}}{1+\alpha_{\rm eff}} \kappa_{\rm sw} \,,
\ee
with the plasma velocity profile $v(\xi)$ being dependent on $v_w$ and $\alpha$, as described in~\cite{Espinosa:2010hh}.

If the sound wave period is {significantly} shorter than a Hubble time, a sizable fraction of the phase transition energy can go into turbulence in the plasma {when the plasma enters the non-linear regime after the acoustic (sound wave) period.} Turbulence also sources GWs, and its contribution to the GW spectrum is given by~\cite{Caprini:2009yp}
\be \label{eq:Omegaturb}
\Omega_{{\rm turb},*} = 
6.8  
(H_*R_*) \left(1-H_*\tau_{\rm sw}\right) \left(\frac{\kappa_{\rm sw} \,\alpha }{1+\alpha }\right)^{3/2} 
\frac{\left(\frac{f_*}{f_{\rm turb}}\right)^3 \left[1+\left(\frac{f_*}{f_{\rm turb}}\right)\right]^{-\frac{11}{3}}}{1+8\pi f_*/H_*}\,,
\ee
with the peak frequency,
\be
f_{\rm turb} = 3.9 ((v_w-c_s)  \, R_*)^{-1} 
\,.
\ee
The factor $1-H_*\tau_{\rm sw}$ in~\eqref{eq:Omegaturb} parametrizes the relative amount of vortical motion induced in the plasma, leading to the generation of turbulence. Here we assume that all the energy left in the bulk fluid motion when the fluid flow becomes non-linear and the sound wave period ends is transferred into turbulence, so our estimate corresponds to an approximate upper bound on the turbulent energy component (besides being transferred into turbulence, the bulk motion from the fluid could be converted into heat during the non-linear fluid regime). Nevertheless, a turbulence GW spectrum of comparable amplitude to that from sound waves, as obtained via~\eqref{eq:Omegaturb}, is to be expected when shocks and non-linearities develop in the plasma within a Hubble time~\cite{Hindmarsh:2017gnf}, i.e, when $\tau_{\rm sw} < 1/H_*$. 

%
%

With Eqs.~\eqref{eq:Omegasw} and \eqref{eq:Omegaturb} as an estimate for the GW spectrum from sound waves and turbulence, we find that the peak is still produced by sound waves, but the GW spectrum {may be} quickly dominated by the turbulent contribution {for off-peak frequencies. We stress that}
while our treatment is clearly very simplified and more study of the turbulence contribution is necessary~\cite{Caprini:2009yp,Gogoberidze:2007an,Niksa:2018ofa,Pol:2019yex}, it indicates that in general turbulence will play a much more important role compared to sound waves than previously thought.

\vspace{1mm}

There is one more possible source of GWs, namely oscillations of the field occurring after collisions of very energetic bubble walls~\cite{Child:2012qg,Cutting:2018tjt}. However, one has to remember that in any phenomenologically viable scenario the energy stored in these oscillations
{eventually} has to be converted to thermal energy to end the effective matter domination period they would induce. In practice, the energy stored in oscillations will be damped exponentially with the decay rate of the field $\rho_{\rm osc}\propto \exp(-t/\Gamma_{\rm dec})$. Thus this {GW} source could only play a role in cases where the reheating is very inefficient and the matter-like evolution persists for nearly a Hubble time. However, even then the amplitude of this contribution will be suppressed by the ratio of the mass scale of our transition (proportional to the vacuum expectation value of the field) to the Planck mass~\cite{Cutting:2018tjt,Fairbairn:2019xog}, which will be very small in all the scenarios we consider. Thus, the only persisting modification to our predictions can come from modified redshifting during the matter-like evolution period, an issue we address in the following Section.

\section{Expansion rate of the Universe}
\label{sec:expansion}

The probability $P$ that a given point still remains in the false vacuum is given by~\cite{Guth:1979bh,Guth:1981uk}
\be
\label{eq:prob_false_vacuum_T}
P(T)\equiv e^{-I(T)}\, , \quad I(T) \equiv \frac{4\pi}{3}\int_{T}^{T_c}\frac{dT' \,\Gamma(T')\, v_w^3}{T'^4 H(T')} \,\left[ \int_{T}^{T'}\frac{d T''}{H(T'')} \right]^3 \,,
\ee
where $v_w$ is the velocity of the wall of the expanding bubble, and
\be
\Gamma(T) \equiv T^4\left(\frac{S_3}{2\pi T}\right)^{\frac{3}{2}} \,e^{-S_3/T}
\ee
is the bubble nucleation rate per unit of time and volume.
Taking into account the fraction of the false vacuum energy already converted into bubble wall energy 
\be
1-P(T) = \int_{T}^{T_c} dT' \frac{d P(T')}{dT'} \, ,
\ee
which redshifts as radiation for $v_w\simeq 1$~\footnote{For smaller $v_w$ the bubble wall energy density scales slower than radiation~\cite{Kolb:1990vq}.}, the Hubble rate during the phase transition is
\be \label{eq:Hubble}
\begin{split}
H^2 &= \frac{1}{3 M_{\mathrm{pl}}^2} \left( \rho_R + \rho_V + \rho_{\rm wall} \right) \\
&= \frac{1}{3 M_{\mathrm{pl}}^2} \left[\frac{\pi^2}{30} g_{\rm eff}(T) T^4 + \Delta V P(T)+ \Delta V \int_{T}^{T_c} dT' \frac{d P(T')}{dT'} \left(\frac{h_{\rm eff}(T) T^3}{h_{\rm eff}(T') T'^3}\right)^{4/3}\right] \,,
\end{split}
\ee
where $g_{\rm eff}(T)$ and $h_{\rm eff}(T)$ are, respectively, the effective number of relativistic energy and entropy degrees of freedom at temperature $T$. 

The standard adiabatic assumptions still hold because, even though the energy stored in the bubble walls, described by the last term in {Eq.~\eqref{eq:Hubble}}, redshifts like radiation, it does not contribute to the energy density of the thermal plasma described by the first term. Thus the amount of primordial radiation just scales as $a^{-4}$, and it still is a good measure of the expansion. The new difficulty we encounter is that the probability of remaining in the false vacuum depends on the expansion rate. However, the coupled Eqs.~\eqref{eq:prob_false_vacuum_T} and \eqref{eq:Hubble} can be solved simply by iterating them starting from~\eqref{eq:Hubble} with $P(T)=1$. The system converges very rapidly. 

Independently of the model, we find (see Section~\ref{sec:examples}) that the effect of the above correction on the percolation temperature is negligible. Moreover, it is a good approximation that vacuum dominance lasts until percolation. After the bubble collisions, the field oscillates in a roughly quadratic potential around the minimum. The energy density of this oscillating scalar field redshifts like matter so, if the decay rate of the scalar is small, $\Gamma_{\rm dec} < H$, the universe can experience an early period of matter dominance until $\Gamma_{\rm dec} \simeq H$ when the oscillating field decays reheating the plasma. Assuming that the transition was strongly supercooled, $\Delta V \gg \rho_\gamma(T_*)$, and that the thermalization after the decay is fast, $a_{\rm dec} = a_{\rm reh}$, the reheating temperature $T_{\rm reh}$ is given by energy conservation, $\rho_\gamma(T_{\rm reh}) = (a_*/a_{\rm dec})^3 \Delta V$, as
\be
T_{\rm reh} = \left(\frac{\Gamma_{\rm dec}}{H_*}\right)^\frac12 \left(\frac{30\Delta V}{\pi^2 g_{\rm eff}(T_{\rm reh})} \right)^\frac14 \,,
\ee
where $H_* = 3 M_P^2 \Delta V^2/(8\pi)$ is the Hubble rate at the percolation temperature. If $\Gamma_{\rm dec} > H_*$, radiation dominance begins right after percolation and the above equation for $T_{\rm reh}$ holds when setting $\Gamma_{\rm dec}/H_* = 1$.

Finally, we calculate the GW signal today. The spectrum given in Section~\ref{sec:gws} is calculated at $T=T_*$, and redshifts as the universe expands. The amplitude of the signal scales as radiation~\footnote{We approximate that $g_{\rm eff}(T_{\rm reh}) = h_{\rm eff}(T_{\rm reh})$.}:
\be
\begin{aligned} \label{eq:Omegaredshift}
\Omega_{{\rm GW},0} &= \left(\frac{a_*}{a_0}\right)^4  \left(\frac{H_*}{H_0}\right)^2 \Omega_{{\rm GW},*} 
=  1.67\times 10^{-5} h^{-2} \left(\frac{100}{g_{\rm eff}(T_{\rm reh})}\right)^\frac13 \left(\frac{\Gamma_{\rm dec}}{H_*}\right)^\frac23 \Omega_{{\rm GW},*} \,,
\end{aligned}
\ee
and the frequency as $f\sim a^{-1}$:
\be
\begin{aligned} \label{eq:fredshift}
f_0 & = \frac{a_*}{a_0} f_* 
= 1.65\times 10^{-5} \,{\rm Hz}\, \left( \frac{T_{\rm reh}}{100\,{\rm GeV}} \right) \left( \frac{g_{\rm eff}(T_{\rm reh})}{100} \right)^\frac{1}{6} \left(\frac{\Gamma_{\rm dec}}{H_*}\right)^{-\frac{1}{3}} \left(\frac{f_*}{H_*}\right) \,.
\end{aligned}
\ee
Again, if $\Gamma_{\rm dec} > H_*$, we should set $\Gamma_{\rm dec}/H_* = 1$ in the above equations. In that case our results match those given in Ref.~\cite{Kamionkowski:1993fg}.

\section{Illustrative examples} 
\label{sec:examples}

We are now ready to apply the formalism discussed in the previous Sections to explore in which models the GW signals can have a significant bubble collision contribution. {In the following we analyse two representative models as a way to illustrate our results for scenarios with polynomial and conformal scalar potentials.} We start with a generic extension of the SM potential by an $|H|^6$ non-renormalisable term in Section~\ref{sec:H6}, and consider subsequently 
{an extension of the SM by a spontaneously broken $U(1)_{B-L}$ gauge symmetry featuring classical scale invariance -- which allows for a prolonged period of supercooling -- in Section~\ref{sec:csi}.}

In each model we compare the produced GW signal to the sensitivities of different phases of LIGO~\cite{TheLIGOScientific:2014jea,Thrane:2013oya,TheLIGOScientific:2016wyq}, as well as the projected experiments LISA~\cite{Bartolo:2016ami}, MAGIS~\cite{Graham:2016plp,Graham:2017pmn} and AION~~\cite{AION:2018}~\footnote{For the km configuration of MAGIS/AION we assume a baseline length of $2$ km with phase noise  $0.3 \times10^{-5}/\sqrt{\rm Hz}$ and large momentum transfer $4\times 10^4$ together with the sensitivity prescription in~\cite{Graham:2016plp}, whereas for the space version we assume the configuration discussed in~\cite{Graham:2017pmn}.}. 
and the
Einstein Telescope~(ET)~\cite{Punturo:2010zz,Hild:2010id}.
We also show the envisioned sensitivities of the proposed future missions DECIGO~\cite{Kawamura:2006up} and Big Bang Observer (BBO)~\cite{Yagi:2011wg,Crowder:2005nr}. 
To quantify the detectability of the signal $\Omega_{\rm GW}(f)$ we calculate the signal-to-noise ratio,
\be
{\rm SNR} \equiv \sqrt{\mathcal{T}\int {\rm d}f\, \left[\frac{\Omega_{\rm GW}(f)}{\Omega_{\rm sens}(f)}\right]^2} \,,
\ee
where $\mathcal{T}$ is the observation time and $\Omega_{\rm sens}(f)$ the sensitivity of the given detector. The signal is {considered to be} detectable if the signal-to-noise ratio is above some threshold value ${\rm SNR}_{\rm thr}$. In the power-law integrated sensitivities~\cite{Thrane:2013oya} shown with the spectra we use $\mathcal{T}=5\,${years} and ${\rm SNR}_{\rm thr}=10$.

\subsection{Standard Model with $\left|H\right|^6$ non-renormalisable term}
\label{sec:H6}

The generation of GWs in this model has been analysed previously in~\cite{Ellis:2018mja}. We nevertheless use it here as a means to illustrate our treatment for the case of polynomial scalar potentials, as well as to obtain the possible GW contribution from bubble collisions in this scenario, discussing also at the end of this Section various improvements in the GW treatment compared to the analysis in~\cite{Ellis:2018mja}.

We start by computing $\alpha_\infty$ and $\alpha_{\rm eq}$ in the Standard Model. Using $(N_W,N_Z,N_t)=(6,3,12)$ in~\eqref{eq:Deltam} and neglecting other contributions that are usually small~\cite{Espinosa:2010hh,Caprini:2015zlo} we compute $\alpha_\infty$ with~\eqref{eq:alpha_infty}. Similarly, using $(N_W,N_Z)=(6,3)$ in~\eqref{eq:Deltam} with $g=g'$ we can compute $\alpha_{\rm eq}$ from~\eqref{eq:alpha_prime_infty}. We obtain
\be \label{eq:SMH6_alphas}
\alpha_\infty = 4.8 \times 10^{-3} \left( \frac{\phi_*}{T_*} \right)^2 \, , \quad 
\alpha_{\rm eq} = 7.3 \times 10^{-4} \left( \frac{\phi_*}{T_*} \right)\, ,
\ee
where $\phi_*$ is the position of the minimum of the potential at temperature $T_*$. These estimates {for $\alpha_\infty$ and $\alpha_{\rm eq}$} also hold in simple extensions of the SM involving extra scalars, as such new particles do not become massive due to the transition and the only {modification compared to the SM} is a slightly higher $g_*$, which is present implicitly through $\rho_R$. {This} cancels, however, in the ratios of $\alpha$s in Eqs.~\eqref{eq:gammaeq2} and \eqref{eq:Ewall} used to compute GW spectra.

\begin{figure}
\begin{center}
\includegraphics[height=0.31\textwidth]{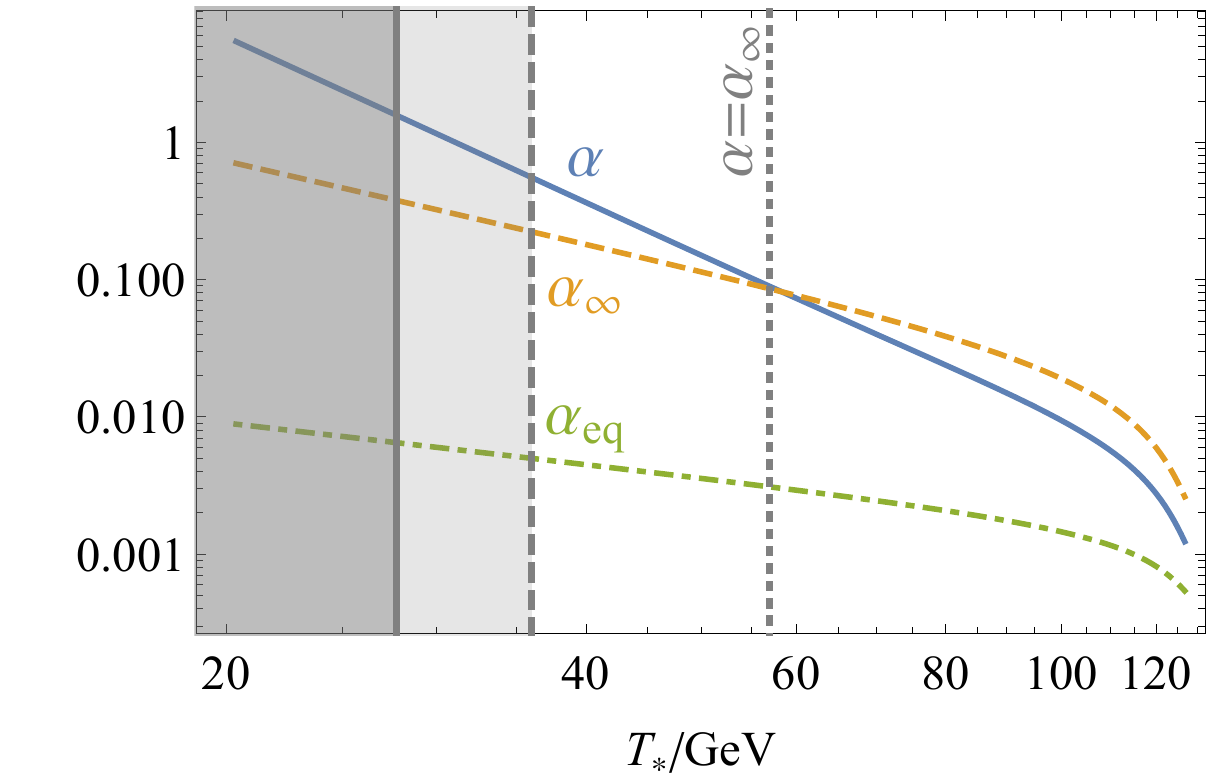} \hspace{1mm}
\includegraphics[height=0.31\textwidth]{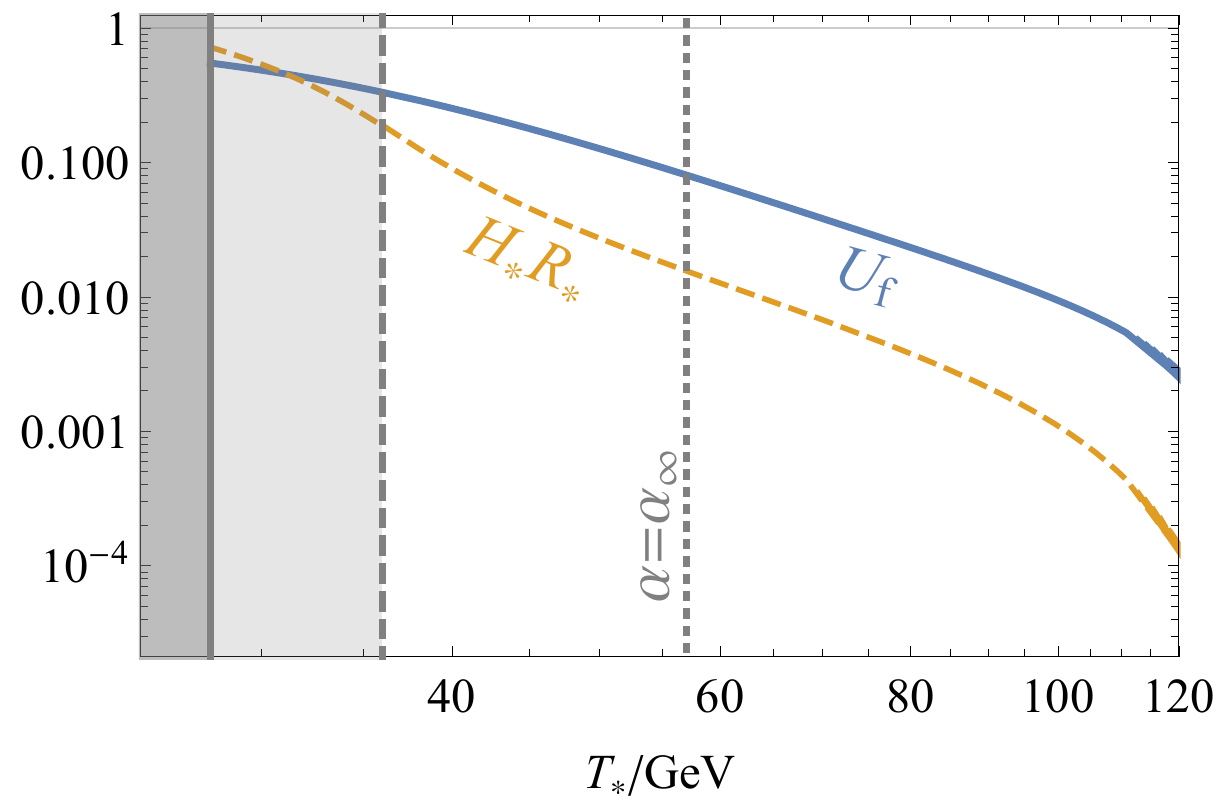}\\ \vspace{3mm}
\hspace{1mm}\includegraphics[height=0.31\textwidth]{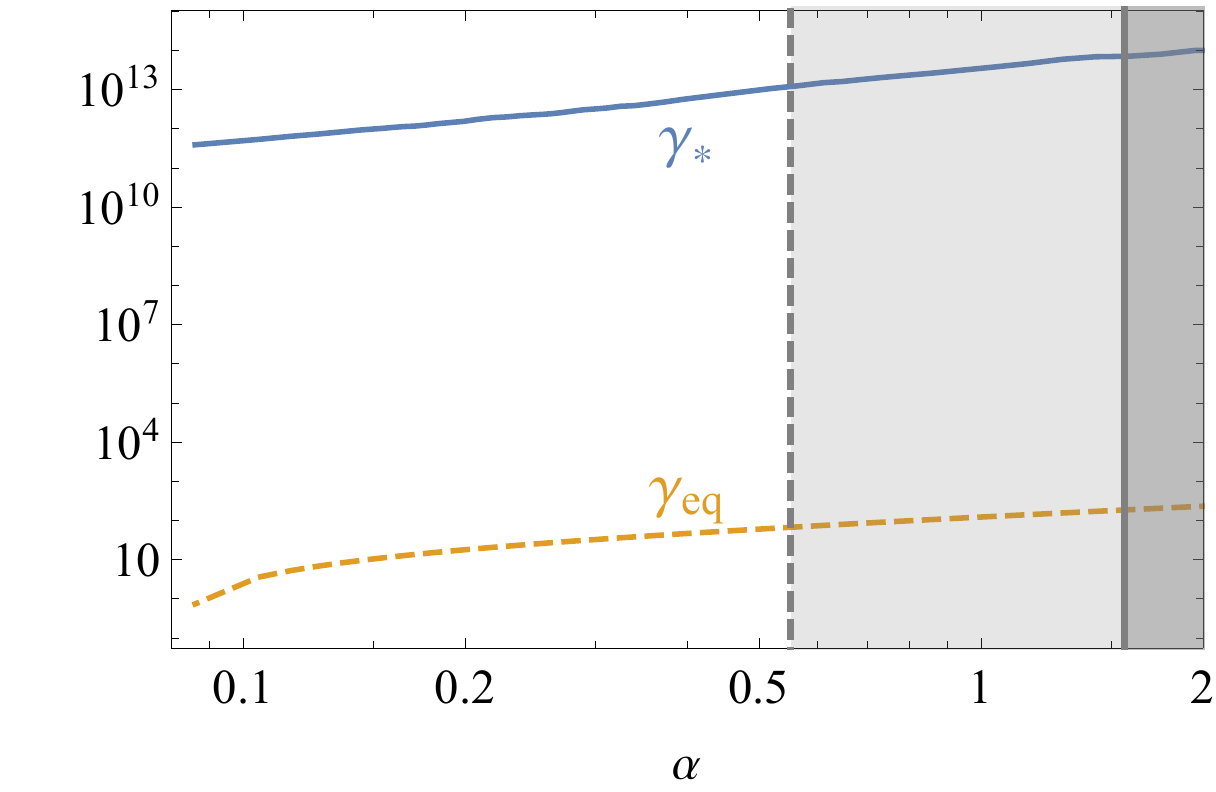} \hspace{1mm}
\includegraphics[height=0.31\textwidth]{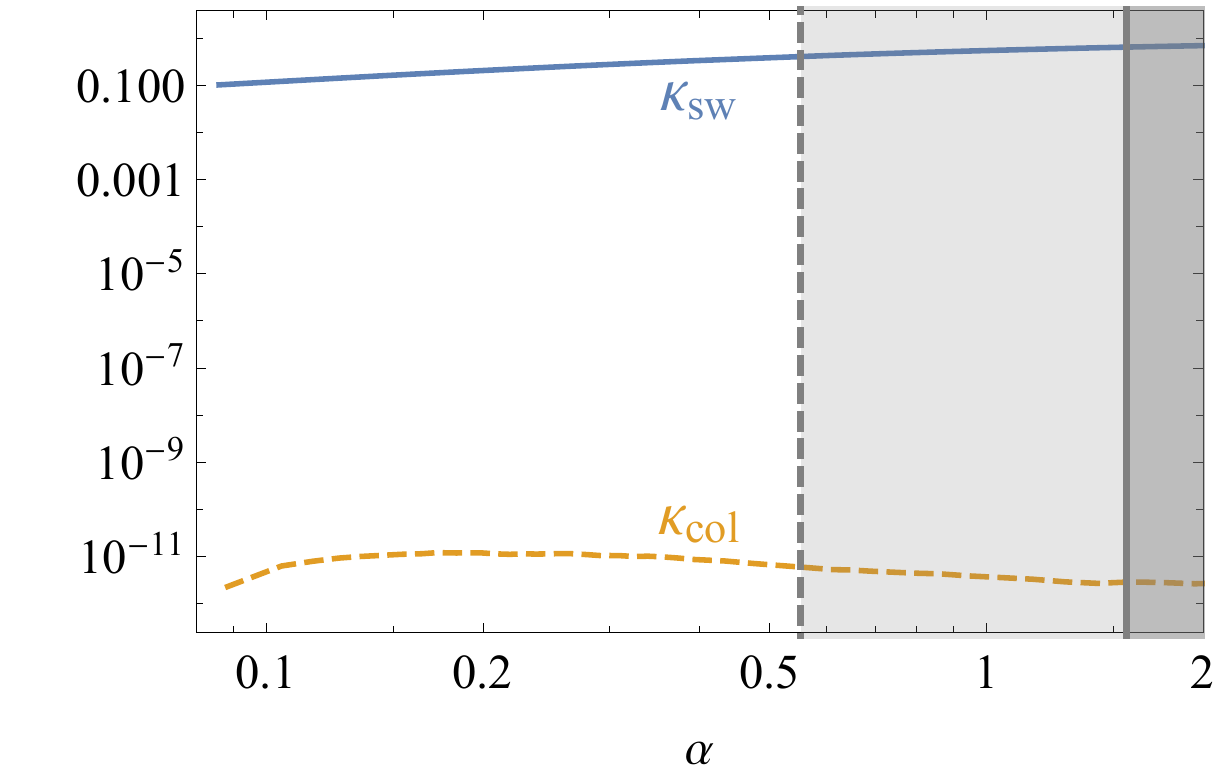}
\caption{ The strength of the transition $\alpha$ in the SM+$H^6$ model as a function of the percolation temperature $T_*$ together with $\alpha_\infty$ and $\alpha_{\rm eq}$ from Eqs.~\eqref{eq:gammaeq2} and~\eqref{eq:gamma_star} (upper left panel) .
 The final size of bubbles at collision $H_* R_*$ and RMS velocity of the plasma $U_f$ used to compute fraction of Hubble time $H_*R_*/U_f$ at which shocks develop in the plasma ending the sound-wave period and starting the turbulence period (upper right panel).
The gamma factor a runaway wall would have,$\gamma_*$, together with the terminal gamma factor the bubbles reach in their expansion $\gamma_{eq}$ (lower left panel).
The resulting efficiency factors for the sound wave GW signal $\kappa_{sw}$ and the bubble collision signal $\kappa_{col}$ (lower right panel). The gray areas in each panel indicates the area of the parameter space excluded because the bubbles do not percolate, while the light gray area separated by a dashed line indicates where percolation is questionable.}
\label{fig:SMH6}
\end{center}
\end{figure}

We perform the standard field profile calculation of the tunneling field solutions for a range of values of the cutoff scale $\Lambda$ associated with the $\left|H\right|^6$ effective operator, from which we find allowed values of $\alpha$ and $R_*$ as functions of the temperature 
{(for details on the scalar potential of the model and the calculation of $\alpha$ and $R_*$, we refer the reader to Ref.~\cite{Ellis:2018mja}).} We show the {values of $\alpha$} together with the corresponding values {of $\alpha_\infty$ and $\alpha_{\rm eq}$} from Eq.~\eqref{eq:SMH6_alphas} in the upper left panel of Fig.~\ref{fig:SMH6}. 
{The upper right panel of the same figure shows the final size of bubbles at collision, $H_* R_*,$ and RMS velocity of the plasma $U_f$. In the region where $H_* R_* < U_f$, shocks/non-linearities develop in the plasma within a Hubble time, ending the sound wave period and potentially starting the turbulence period.
As we see, all reliable (from the point of view of percolation) results are in this region. While a reliable description of processes taking place in the plasma from the termination of the sound wave period is still lacking, it is clear that one cannot simply rely on existing sound-wave simulations to compute the efficiency of the plasma contribution.} The lower {left panel of Fig.~\ref{fig:SMH6} shows} the gamma factor at which bubbles stop accelerating, $\gamma_{\rm eq}$, and the gamma factor {that a bubble accelerating all the way to collision} would have using Eqs.~\eqref{eq:gammaeq2} and~\eqref{eq:gamma_star}. As expected, we find that in this class of models the bubbles reach their {terminal} velocity soon after nucleation and, as a result, most of the energy is transferred into the plasma rather than the bubble wall. This is {apparent from} the lower right panel of Fig.~\ref{fig:SMH6}, which shows the final efficiencies for the sound wave and bubble collision signals.

{Fig.~\ref{fig:SMGWs} shows the GW spectra calculated as outlined in Section~\ref{sec:gws}, {together with the sensitivities of LIGO, LISA, AION/MAGIS and ET. Compared to previous studies (see e.g.~\cite{Ellis:2018mja}), we note} the much stronger turbulence signal coming from taking into account the increased efficiency of turbulent production in cases where the flow of plasma becomes non-linear within a Hubble time, see eq.~\eqref{eq:Omegaturb} and~\eqref{eq:Omegaturb}. Finally, we quantify the reaches of planned detectors by calculating their SNR, as shown in Fig.~\ref{fig:SMSNR}. 

\begin{figure}
\begin{center}
\includegraphics[width=0.48\textwidth]{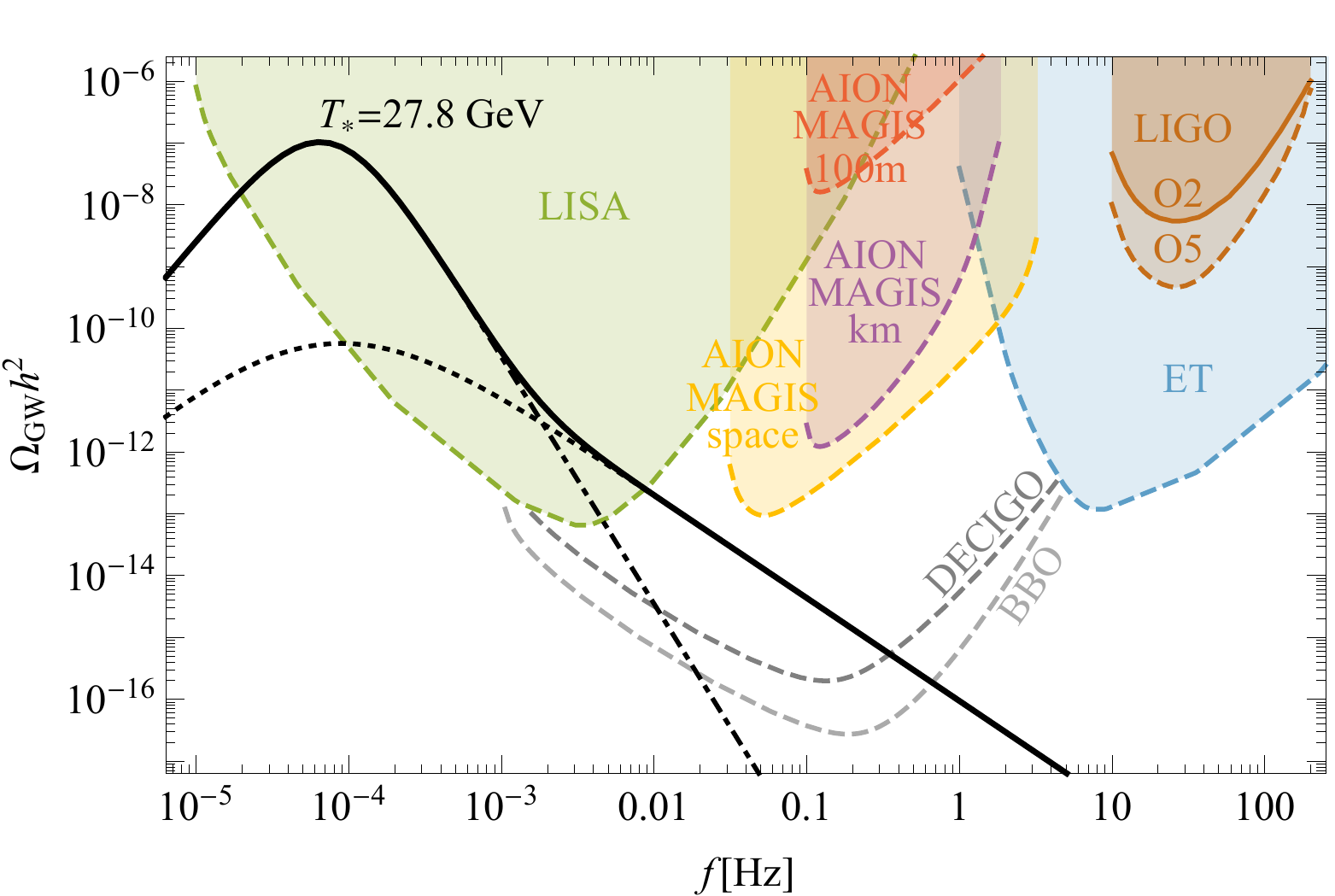} \hspace{2mm}
\includegraphics[width=0.48\textwidth]{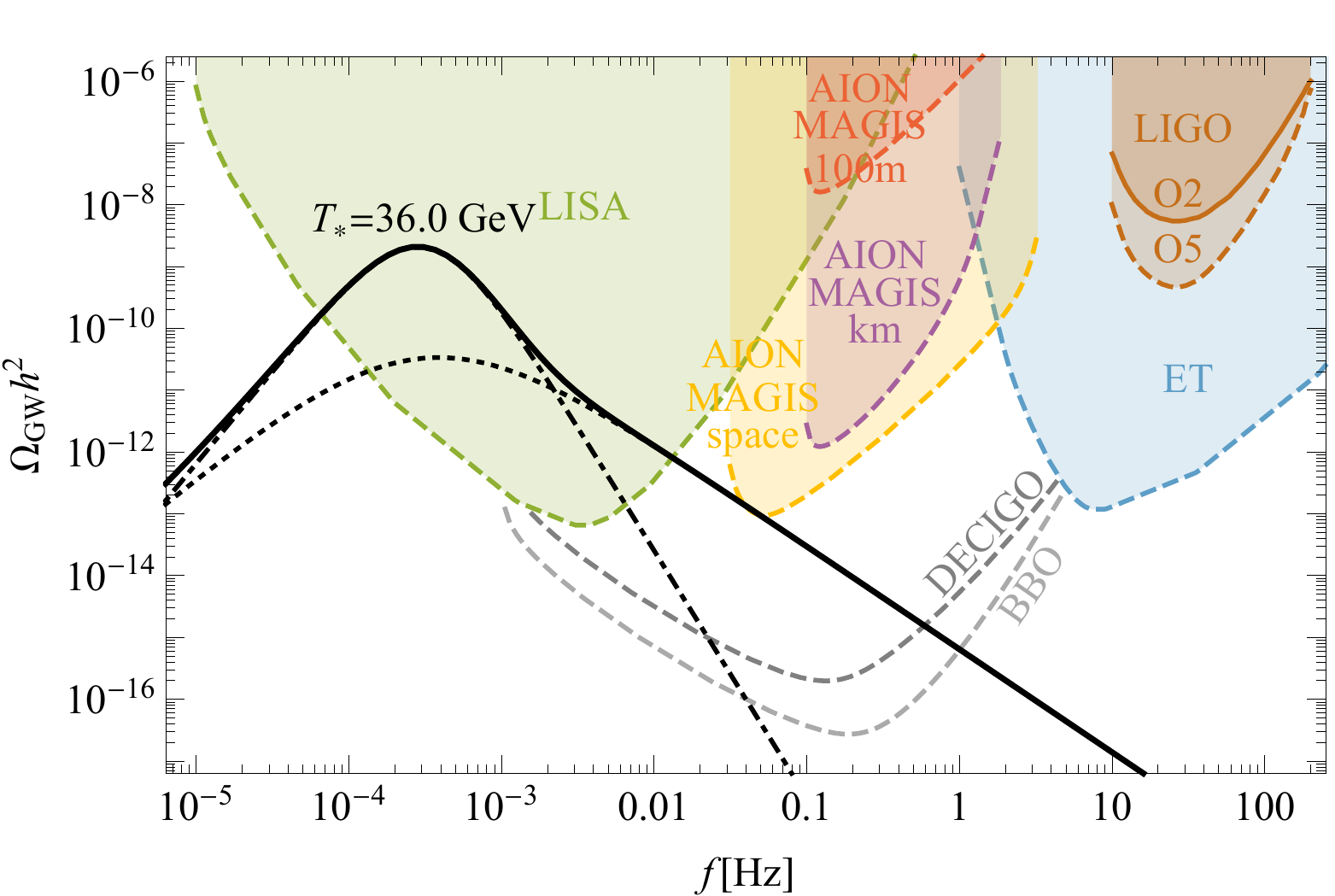}
\caption{{The GW spectra for representative points in the parameter space of the $SM+H^6$ model. The left-hand panel shows the strongest signal not excluded by lack of percolation, while the strongest signal for which percolation is assured is shown on the right-hand panel. The solid line shows the total signal. The dot-dashed and dotted lines show separately the contributions from sound waves and turbulence. The colored lines and regions show the power-law integrated sensitivities of various current and future detectors.}}
\label{fig:SMGWs}
\end{center}
\end{figure}

\begin{figure}
\begin{center}
\includegraphics[height=0.42\textwidth]{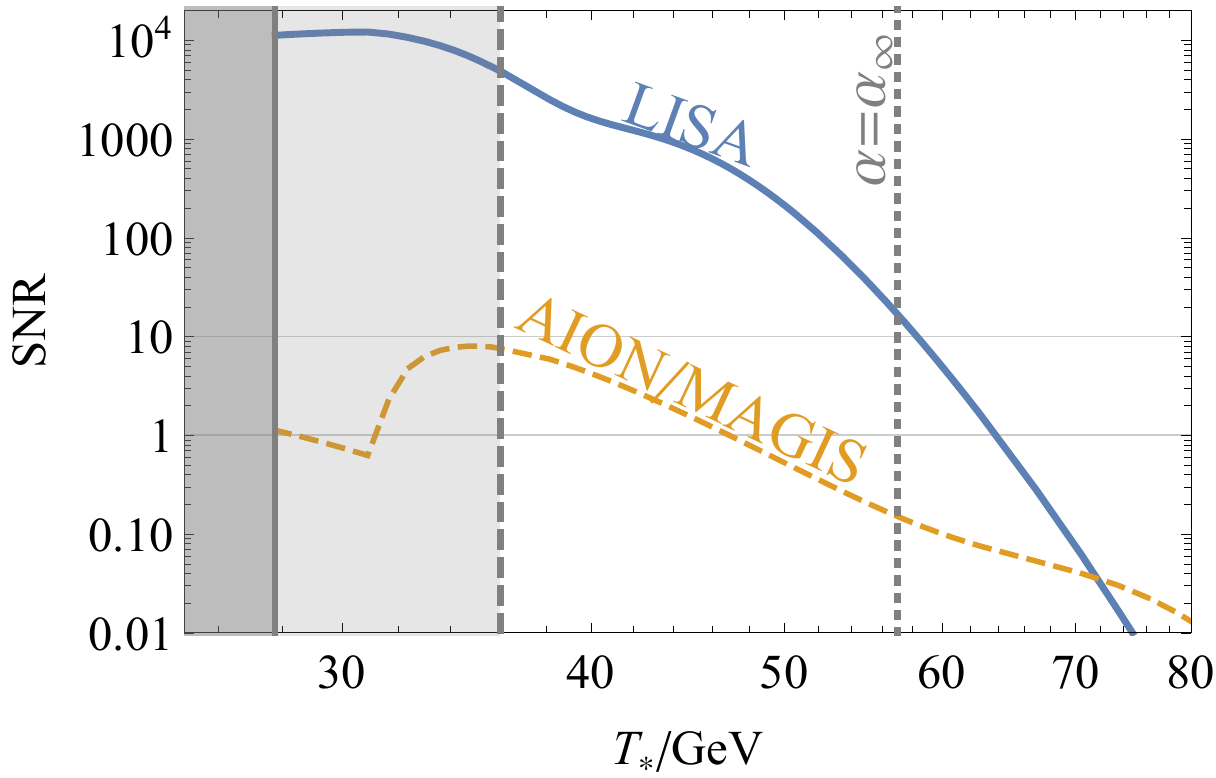} \hspace{2mm}
\caption{{The signal-to-noise ratio (SNR) of the GW signals in the $SM+H^6$ model as could be observed in the planned LISA and AION/MAGIS experiments.}}
\label{fig:SMSNR}
\end{center}
\end{figure}

These quantitative results confirm the usual assumption that in this class of models the bubble collision signal is negligible. 
It is clear that producing a significant bubble collision signal requires a long period of supercooling, which is not allowed here as accelerated expansion would spoil percolation. Interestingly, the maximal supercooling $\alpha\approx 1$ here actually corresponds to a minimum of the collision signal. It increases as the strength of the transition lowers as then the bubbles grow for a shorter time and transfer less energy into the plasma. In fact, in the limit of a very weak transition in which bubbles collide almost immediately after nucleation, the collision efficiency would also be significant. However, in that limit the total energy going into GWs is negligible.    

\begin{figure}
\begin{center}
\includegraphics[height=0.31\textwidth]{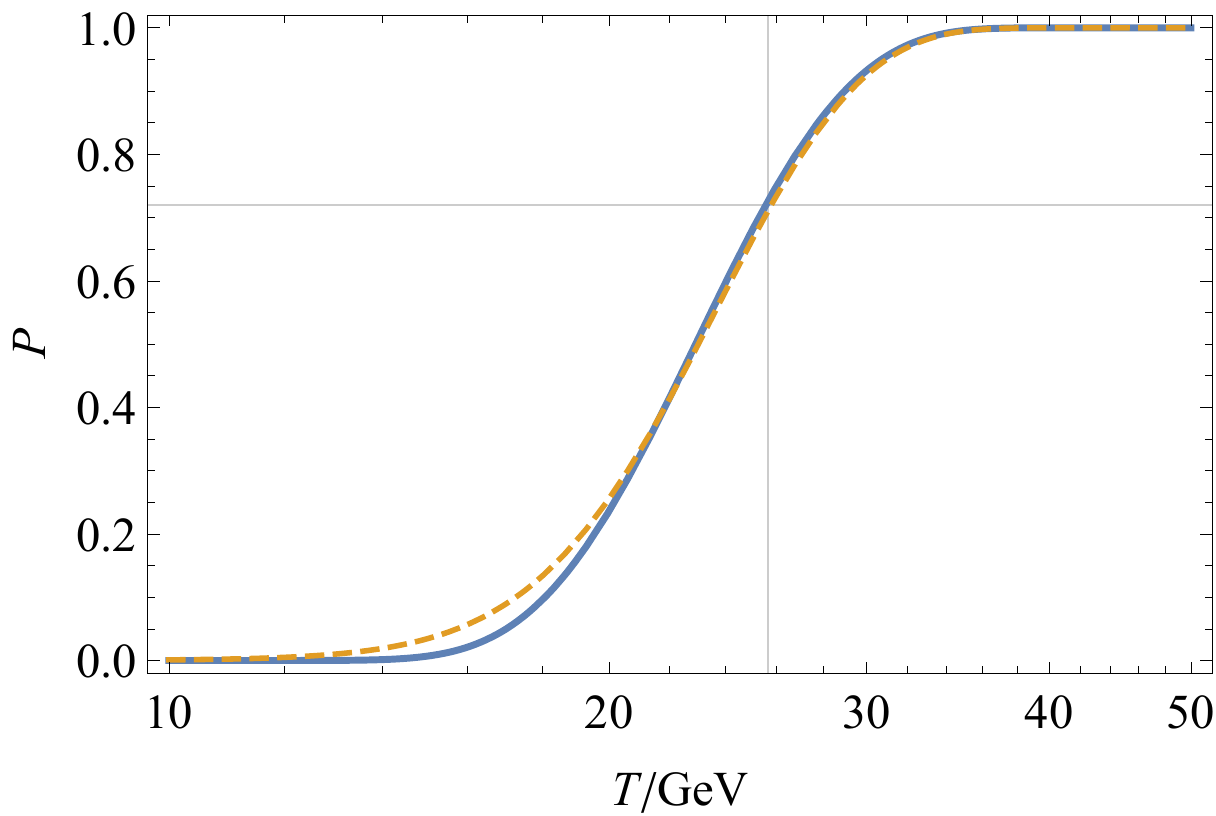} \hspace{3mm}
\includegraphics[height=0.31\textwidth]{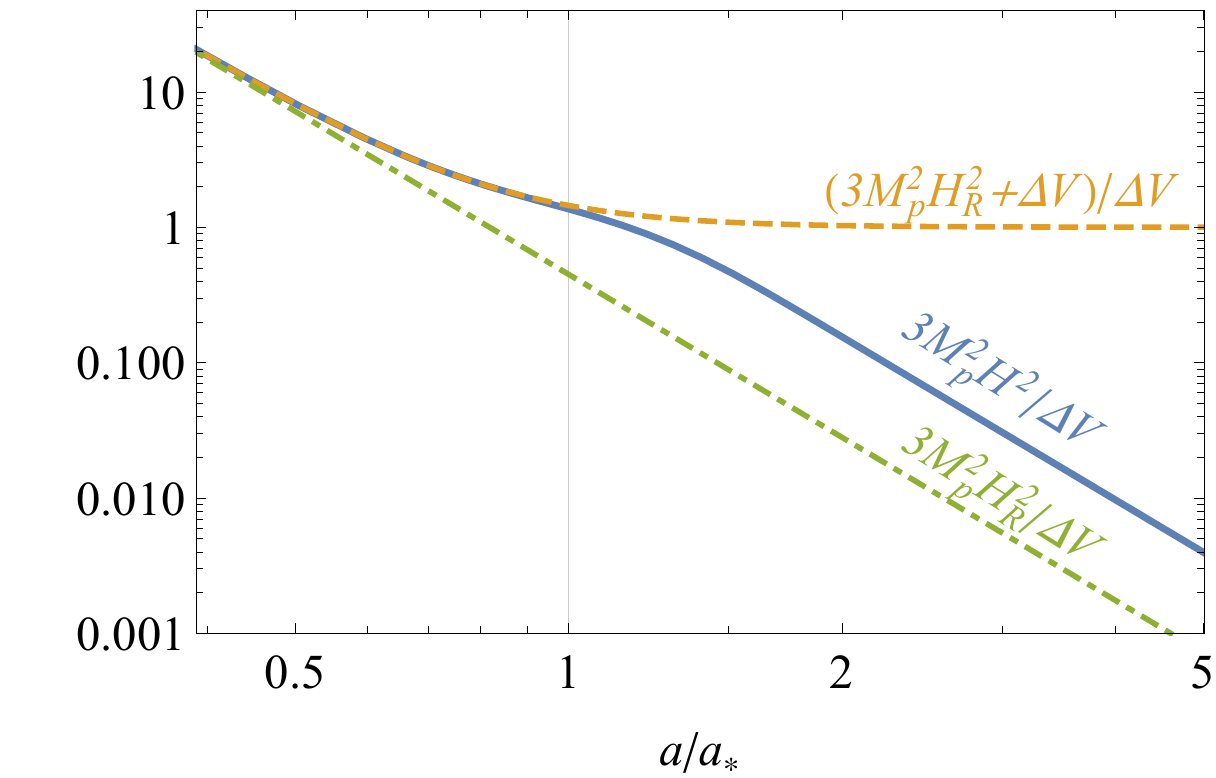} \\ \vspace{2mm}
\includegraphics[height=0.3\textwidth]{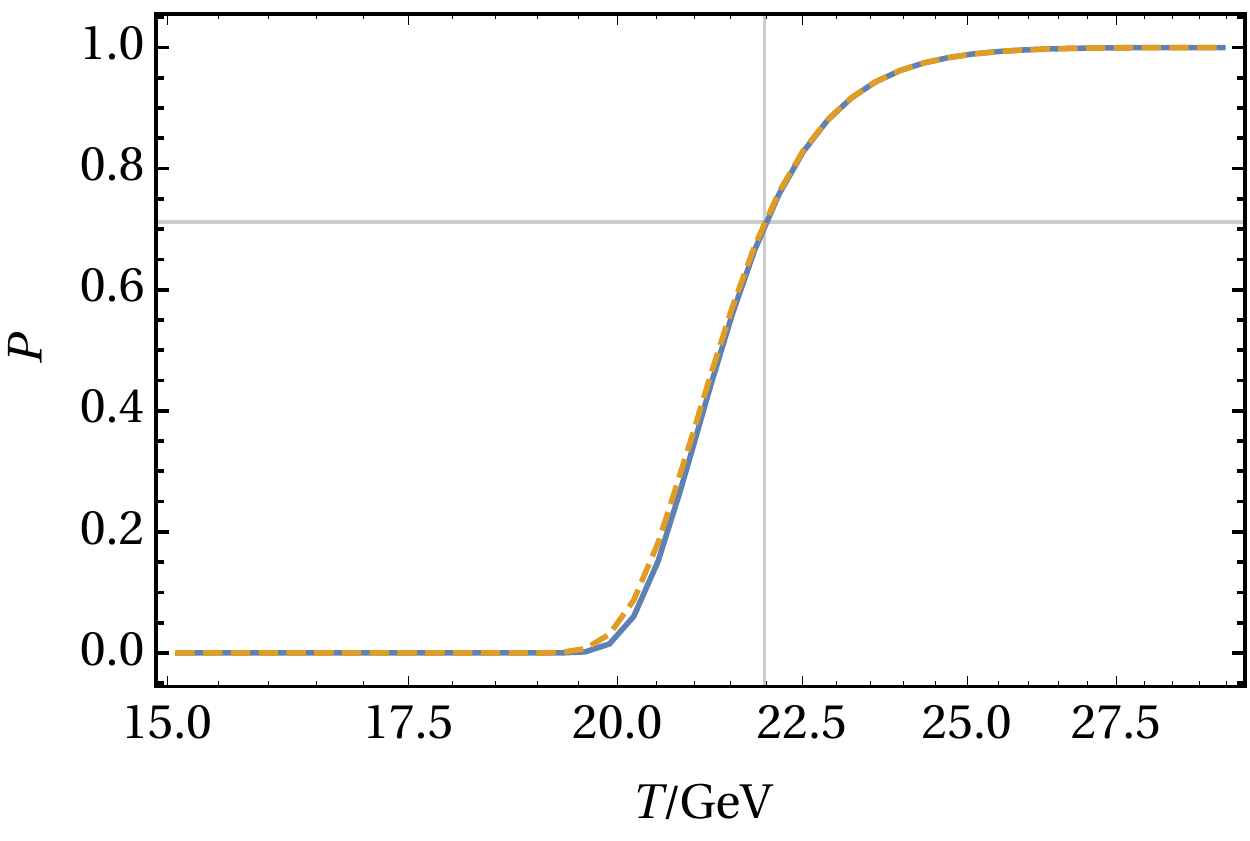} \hspace{11mm} 
\includegraphics[height=0.3\textwidth]{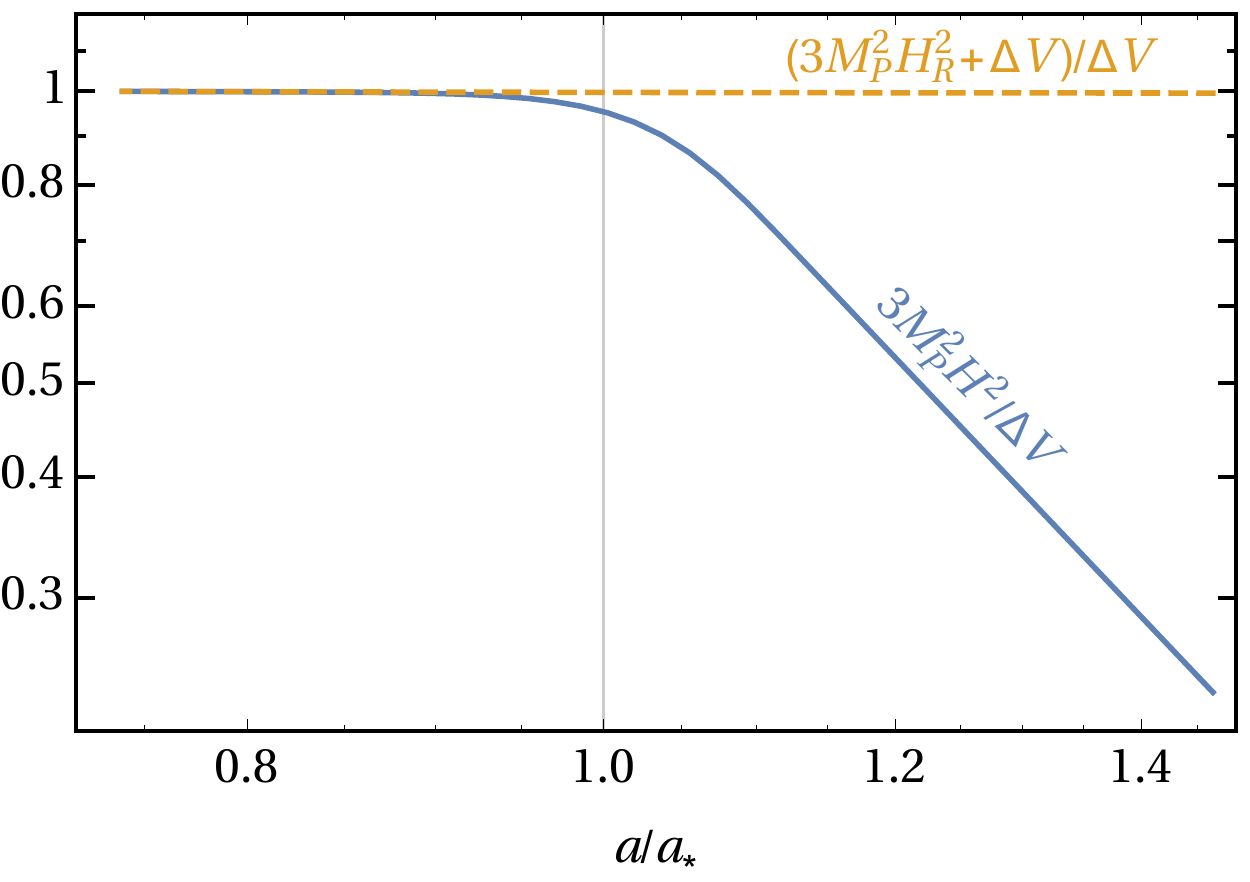}
\caption{The false vacuum probability as a function of temperature (left panels) and the energy density normalized to vacuum energy density as a function of the scale factor (right panels). The onset of percolation is highlighted by the vertical line. The upper panel correspond to the SM with $H^6$ term and the lower panels to the classically scale-invariant model. The yellow dashed line shows the simplest approximation obtained by setting $P=1$ in Eq.~\eqref{eq:Hubble}, and the solid blue one the final result obtained by solving Eqs.~\eqref{eq:Hubble} and~\eqref{eq:prob_false_vacuum_T}, while the green dot-dashed line represents the Hubble rate one would obtain including only the pre-existing thermal background.
}
\label{fig:expansion}
\end{center}
\end{figure}

Before concluding our analysis of this model, we highlight the improved treatment of the false vacuum probability compared to that from~\cite{Ellis:2018mja} as described in Section~\ref{sec:expansion}, taking accurately into account the amount of false vacuum energy converted into bubble-wall energy that redshifts as radiation. The modified probability is shown in the left panel of Fig.~\ref{fig:expansion} and, as we can see, the correction arising from this improved treatment is very small, and does not modify the percolation temperature. The right panel of that figure shows the evolution of the energy density, which is much more realistic with the improved treatment. However, this again influences the results only below the percolation temperature.

\subsection{{Classically conformal $U(1)_{{\rm B}-{\rm L}}$ model}}
\label{sec:csi}

We now consider a classically scale-invariant $U(1)_{{\rm B}-{\rm L}}$ extension of the SM as an example of (classicaly) conformal dynamics. The details of the model can be found in Ref.~\cite{Marzo:2018nov}, and in the following we repeat that calculation except for the GW signal, for which we use the results obtained in the previous section.

The model includes a new scalar field $\varphi$ that couples to the Higgs field $h$ via a quartic portal coupling, $-\lambda_p h^2 \varphi^2/4$. The phase transition first proceeds along the $\varphi$ direction, and is of first order as the thermal corrections induce a barrier between the B$-$L-symmetric and B$-$L-breaking minima. The vacuum expectation value of $\varphi$ then induces a negative mass term for the Higgs field via the portal coupling, that triggers electroweak symmetry breaking. In the following we study the first {step, corresponding to the B$-$L symmetry breaking.} The loop corrections of the effective finite-temperature potential along the $\varphi$ direction are dominated by the B$-$L gauge boson $Z'$, so the dynamics of the B$-$L breaking transition {is} determined mainly by the B$-$L gauge coupling $g_{{\rm B}-{\rm L}}$ and the B$-$L gauge boson mass $m_{Z'}$. Very roughly, the effective finite-temperature scalar potential along $\varphi$ can be approximated as
\be
V(\varphi) \simeq \frac{3 g_{{\rm B}-{\rm L}}^4 \varphi^4}{4\pi^2}\left[\log\left(\frac{\varphi^2}{v_\varphi^2}\right) - \frac{1}{2} \right] \quad +  g_{{\rm B}-{\rm L}} ^2 T^{\,2} \varphi^2 \,,
\ee
where $v_\varphi$ is the vacuum expectation value of $\varphi$. {In all our numerical calculations, following Ref.~\cite{Marzo:2018nov}, we nevertheless use the full} one-loop RG-improved $T=0$ potential, evaluate numerically the thermal $J_T$ functions, and include the resummation of daisy diagrams. As in Ref.~\cite{Marzo:2018nov}, we fix the kinetic mixing parameter to the value $\tilde g = -0.5$.

First, we evaluate the false vacuum probability as described in Section~\ref{sec:expansion}, taking into account the fraction of the false vacuum energy converted into bubble-wall energy. As shown in the lower left panel of Fig.~\ref{fig:expansion}, the correction arising from the latter is very small, and does not change significantly the percolation temperature $T_*$, defined via $P(T_*) = e^{-0.34}$. Hence, in the following results we neglect this, and calculate the percolation temperature simply by setting $P(T)=1$ in Eq.~\eqref{eq:Hubble}, as this makes the numerical evaluation faster. From the lower right panel of Fig.~\ref{fig:expansion} we see that the vacuum dominance lasts roughly until the percolation, after which the energy density of the universe starts to scale as radiation, assuming instantaneous reheating.

The percolation temperature $T_*$ is shown by the blue line as a function of $g_{{\rm B}-{\rm L}}$ in the top left panel of Fig.~\ref{fig:csigamma} for $m_{\rm Z'} = 10\,{\rm TeV}$. The percolation temperature decreases for decreasing values of $g_{{\rm B}-{\rm L}}$, and to the left of the $\alpha=1$ line the transition occurs during vacuum dominance. Eventually the percolation temperature becomes smaller than $T_{\rm QCD}\sim 0.1\,{\rm GeV}$. As first pointed out in Ref.~\cite{Iso:2017uuu} and recently studied in detail in Ref.~\cite{Marzo:2018nov}, the QCD phase transition then changes the dynamics of the B$-$L breaking~\footnote{{See~\cite{vonHarling:2017yew} for an analogous discussion of the effect of the QCD phase transition in the context of a different conformal scenario.}}, as the Higgs field acquires a small vacuum expectation value, $v_{\rm QCD} \simeq 0.1\,{\rm GeV}$, that induces a negative mass term, $-\lambda_p v_{\rm QCD}^2 \varphi^2/4$, for $\varphi$. After the phase transition is completed, the plasma is reheated to a temperature $T_{\rm reh}$, shown by the yellow line. This is calculated assuming instant reheating after percolation. We have checked that, for the parameters used here, the decay rate of $\varphi$,
\be
\Gamma_{\rm dec} \simeq \frac{\lambda_{\rm p}^2 v_\varphi^2}{32\pi m_\varphi} \simeq 2.5\times 10^{-5} g_{{\rm B}-{\rm L}} \left(\frac{m_{Z'}}{10\,{\rm TeV}}\right)^{-3}\,{\rm GeV} \,,
\ee
is larger than the Hubble rate at percolation.

\begin{figure}
\begin{center}
\includegraphics[height=0.31\textwidth]{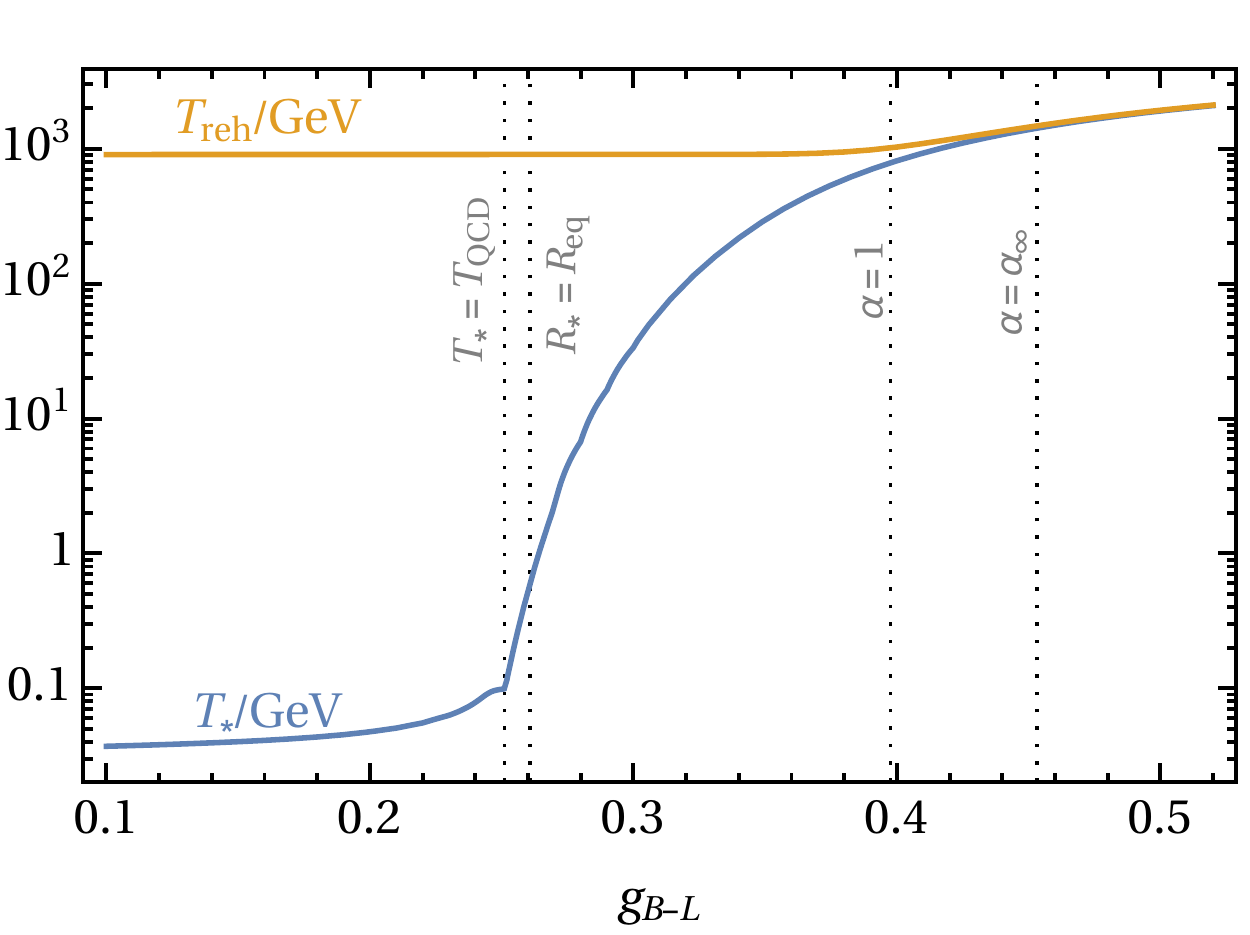} \hspace{4mm}
\includegraphics[height=0.31\textwidth]{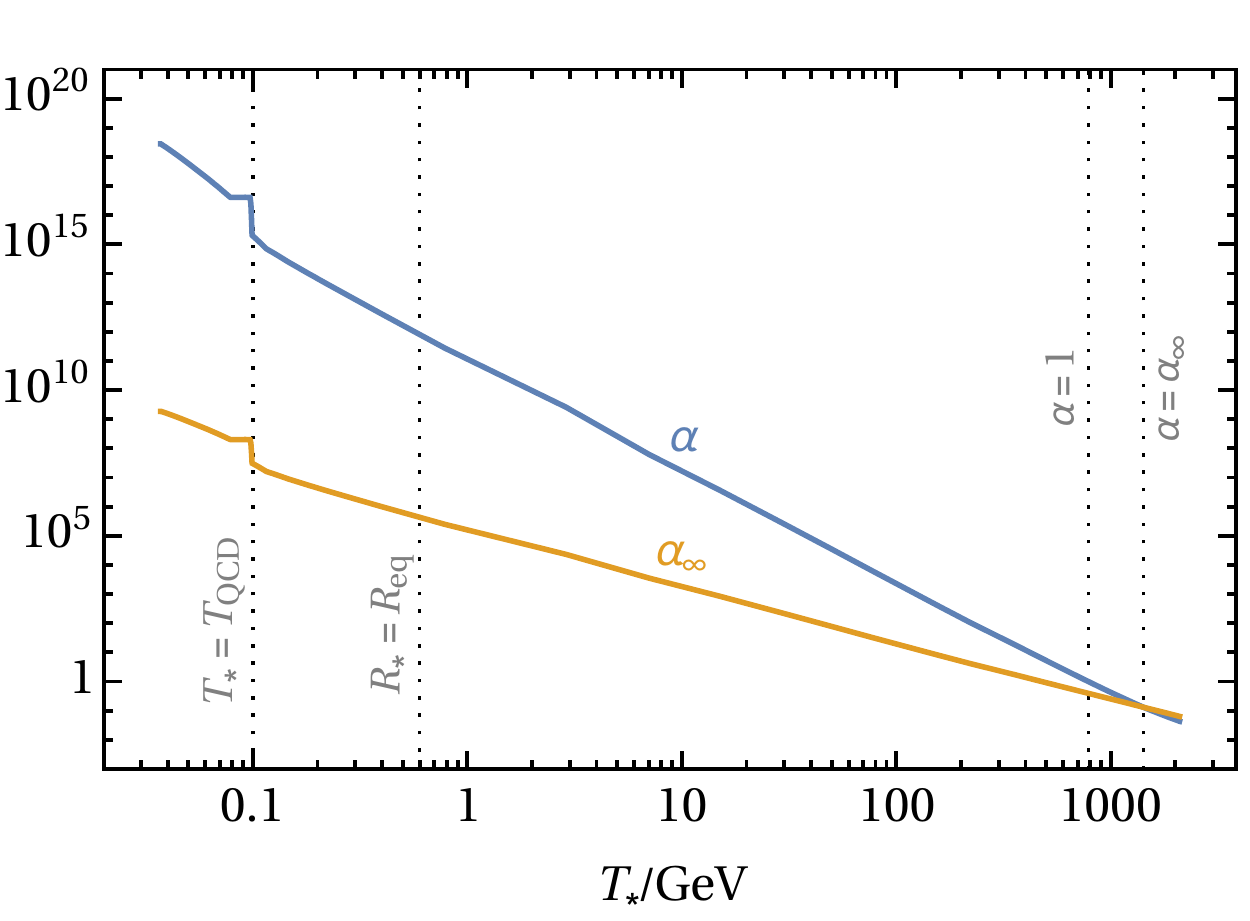} \\ \vspace{1mm}
\includegraphics[height=0.31\textwidth]{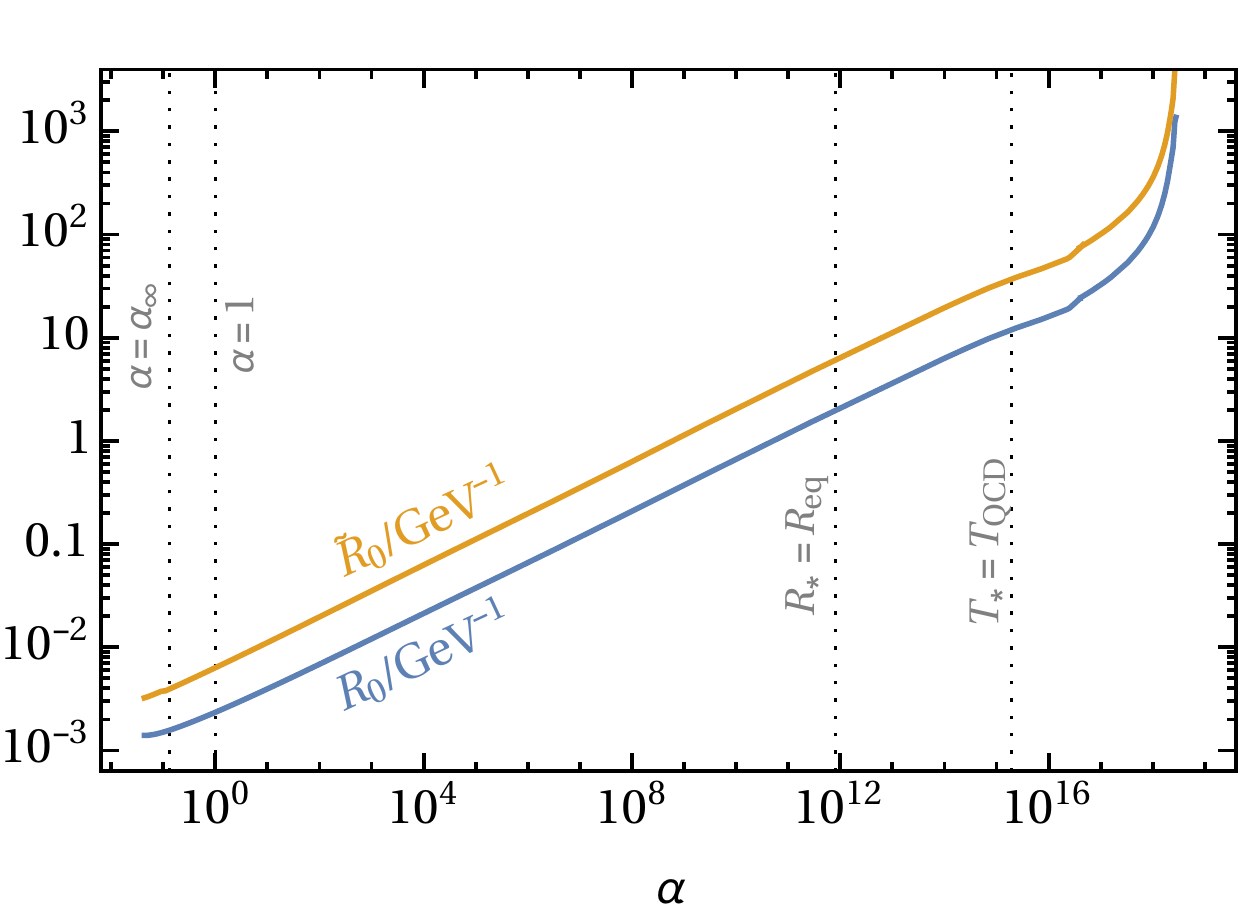} \hspace{4mm}
\includegraphics[height=0.31\textwidth]{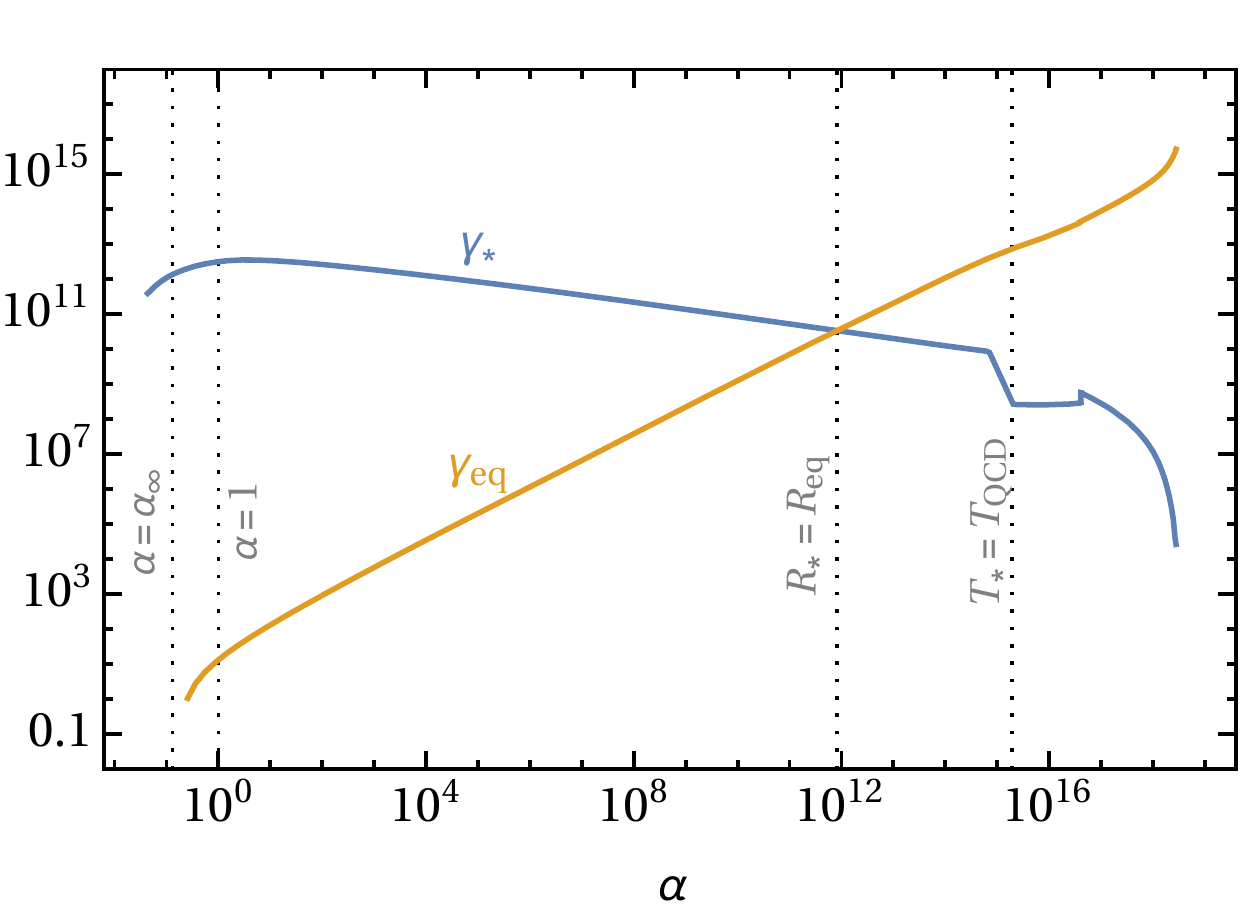} \\ \vspace{1mm}
\includegraphics[height=0.31\textwidth]{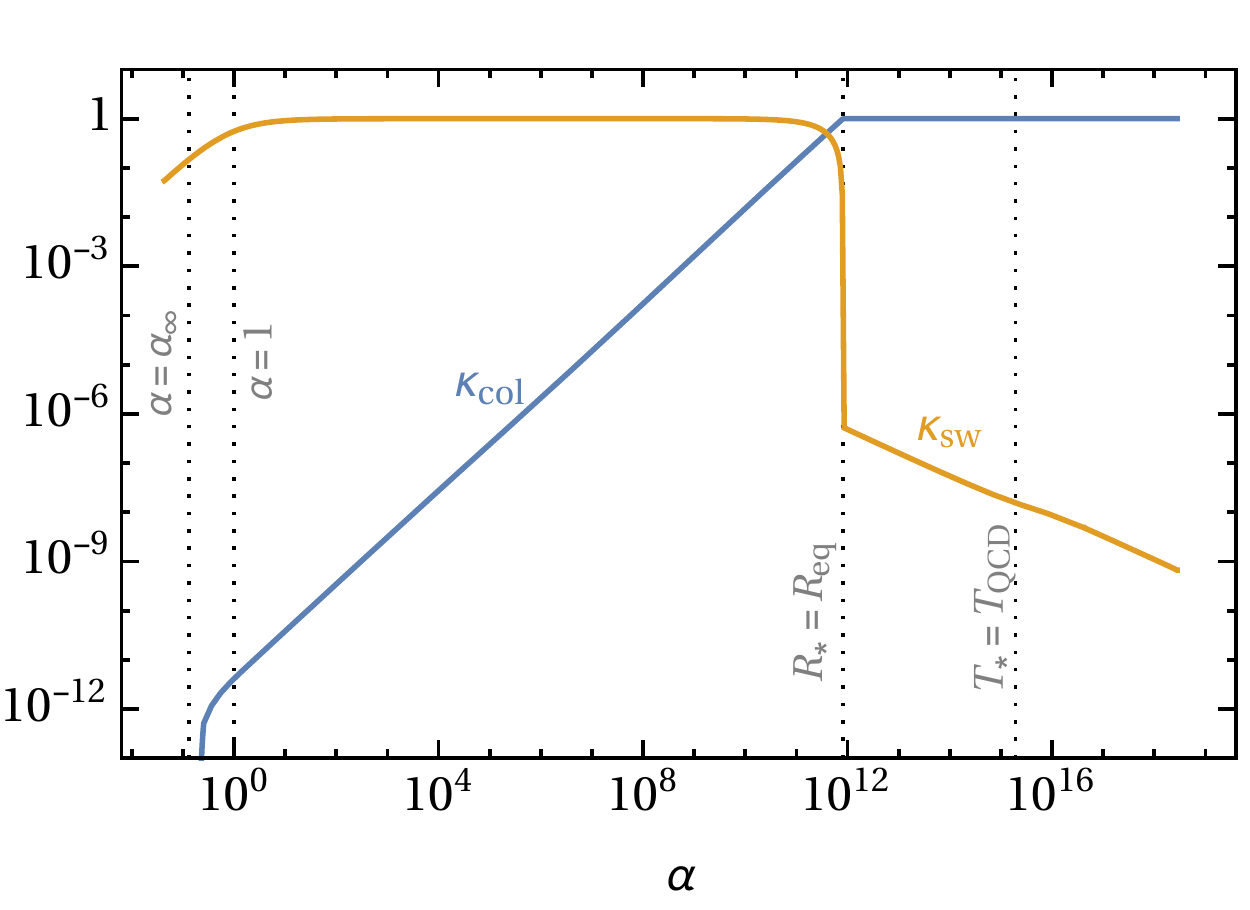} \hspace{4mm}
\includegraphics[height=0.31\textwidth]{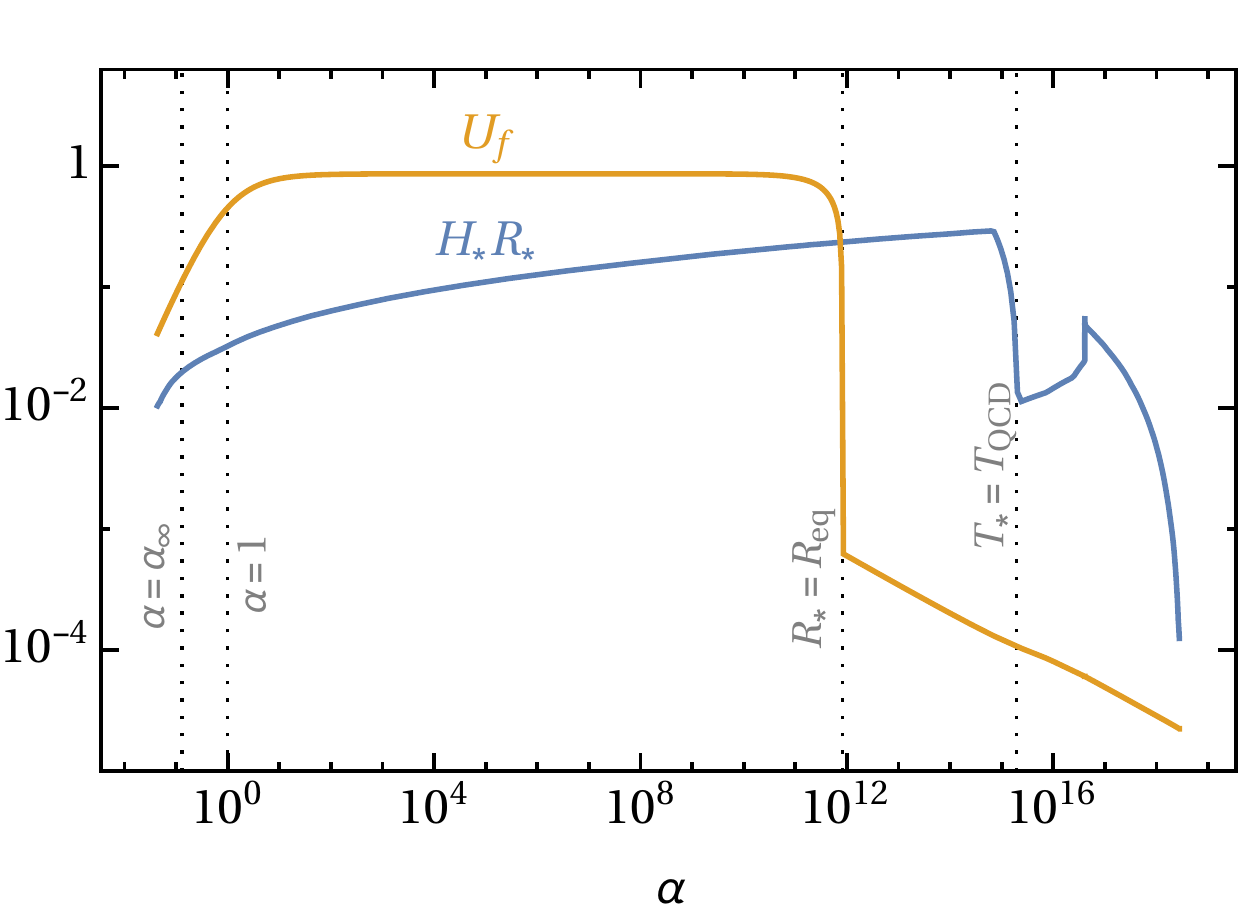}
\caption{ Parameters related to the phase transition in the classically conformal $U(1)_{{\rm B}-{\rm L}}$ model obtained by varying the B$-$L gauge coupling in the range $0.1<g_{{\rm B}-{\rm L}}<0.52$ for fixed $m_{Z'}=10\,{\rm TeV}$.}
\label{fig:csigamma}
\end{center}
\end{figure}

The strength of the transition, characterized by the $\alpha$ parameter, is shown in the top right panel of Fig.~\ref{fig:csigamma}. In the middle left panel the blue line shows the initial bubble radius $R_0$ (in GeV$^{-1}$) calculated from the initial energy, as in Eq.~\eqref{eq:R0}, and used in our analysis. We find that for strong transitions that finish above the QCD scale $R_0$ is roughly given by $1/T_*$.
For the sake of comparison, the yellow line shows the initial radius $\tilde R_0$ determined from the bubble profile, as the radius that maximizes the potential energy. This approximation overestimates the initial radius by a factor of $\sim 3$.

The middle right panel of Fig.~\ref{fig:csigamma} shows the Lorentz $\gamma$ factor of the bubble wall at $T=T_*$, that is given by $\gamma = \min(\gamma_{\rm eq},\gamma_*)$. The bubble wall reaches terminal velocity before percolation for values of $\alpha$ to the left of the $R_*=R_{\rm eq}$ line. As shown in the bottom left panel, the efficiency coefficient for the sound wave contribution on the GW spectrum is in that case close to unity. For larger values of $\alpha$ the terminal velocity is not reached, and the efficiency coefficient for the contribution from bubble collisions becomes almost one. Finally, the RMS fluid velocity is shown in the bottom right panel of Fig.~\ref{fig:csigamma}.

The resulting GW spectrum is shown in Fig.~\ref{fig:csigws} for different values of $g_{{\rm B}-{\rm L}}$, {and $m_{\rm Z'}=10\,$TeV ($m_{\rm Z'}=100\,$TeV) in the left (right) panels. As expected from the efficiency factors shown in the bottom left panel of Fig.~\ref{fig:csigamma}, sound waves are the dominant source of GWs at high percolation temperature, and bubble collisions at low percolation temperature. These correspond to the cases shown in the bottom and top panels of Fig.~\ref{fig:csigws}, respectively. In the intermediate case, which for $m_{Z'} = 10\,$TeV corresponds to $T_*\simeq 1\,$GeV (or $T_*/T_c\simeq3\times 10^{-4}$, where $T_c$ is the critical temperature) and for $m_{Z'} = 100\,$TeV to $T_*\simeq 100\,$GeV (or $T_*/T_c\simeq3\times 10^{-3}$), the GW spectrum has two peaks, as the contributions to the GW signal from bubble collisions and sound waves in the plasma are separated and comparable in amplitude.}

\begin{figure}
\begin{center}
\includegraphics[width=0.48\textwidth]{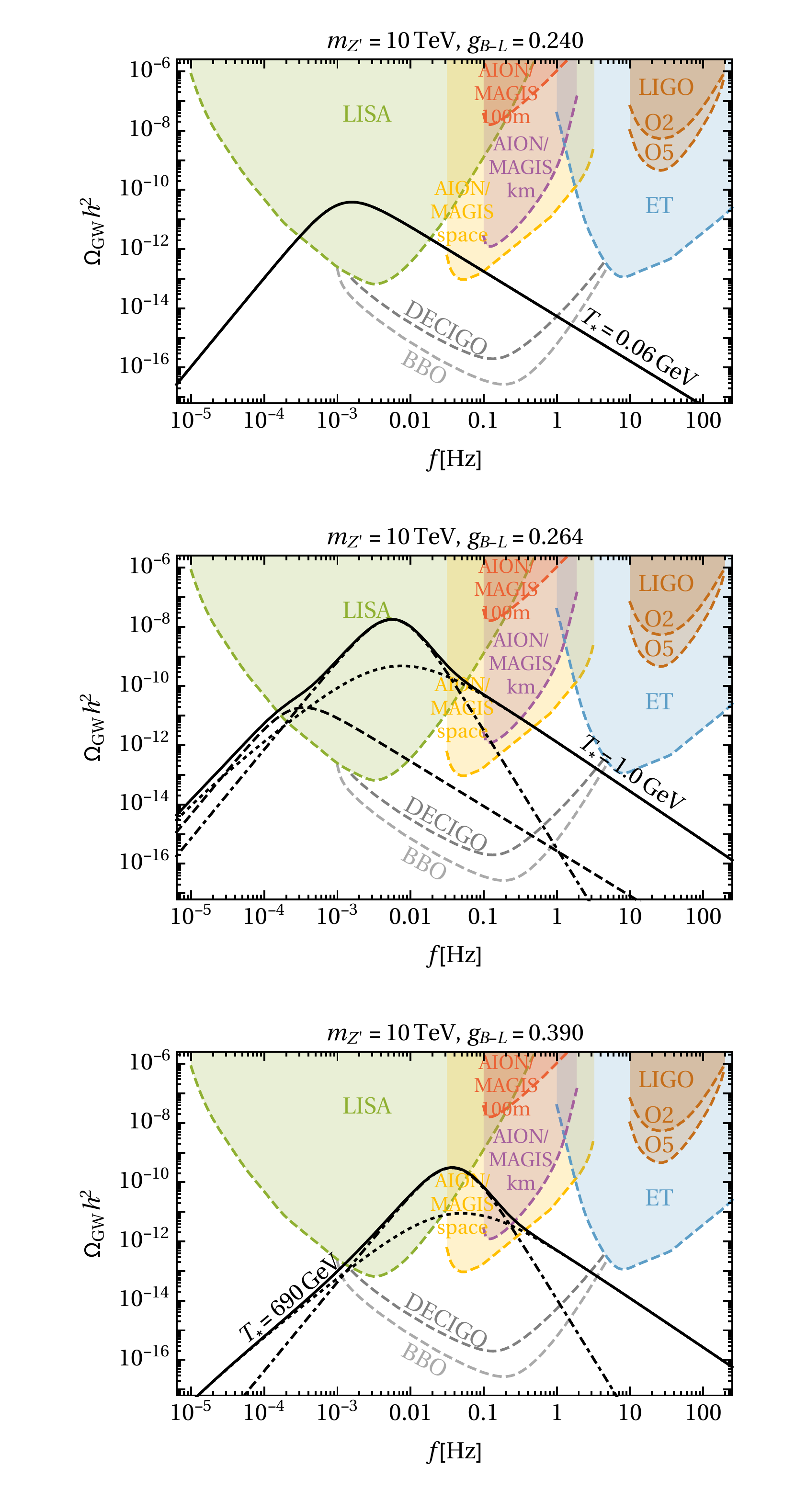} \hspace{1mm}
\includegraphics[width=0.48\textwidth]{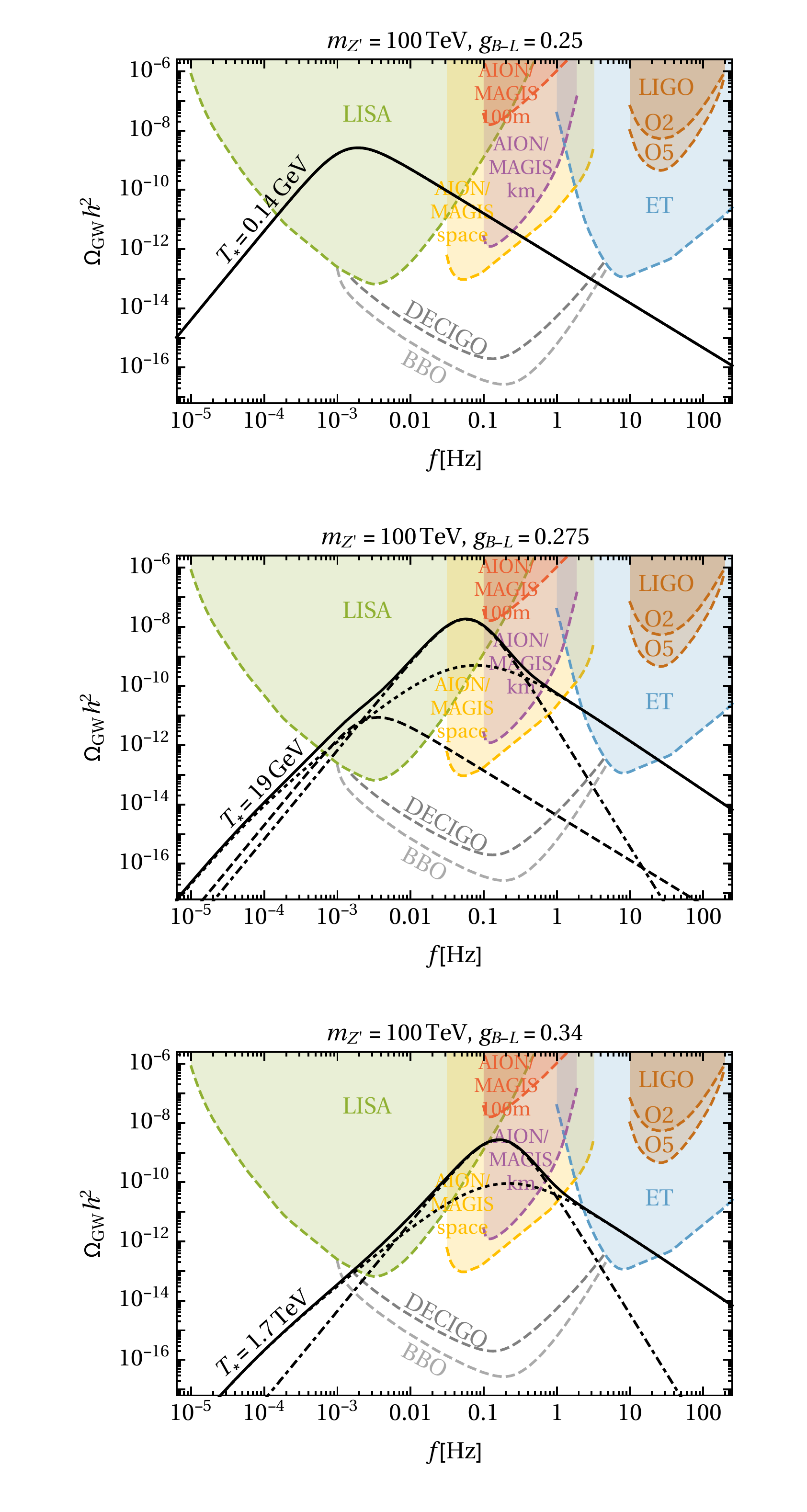}
\caption{The GW signal from the B$-$L breaking phase transition for {$m_{Z'}=10\,{\rm TeV}$ (left panels) and $m_{Z'}=100\,{\rm TeV}$ (right panels), and three different values of $g_{{\rm B}-{\rm L}}$ for both.} The solid line shows the total signal. The dashed, dot-dashed and dotted lines show separately the contributions from bubble collision, sound waves and turbulence. {The colored lines and regions show the power-law integrated sensitivities of various current and future detectors.}}
\label{fig:csigws}
\end{center}
\end{figure}

We also show in Fig.~\ref{fig:csigws} the projected sensitivities of various GW detectors. {We see that in the case of a very strongly supercooled transition the peak of the GW signal from the B$-$L-breaking phase transition generally occurs within the sensitivity of LISA. However, the broadness of the GW spectrum extends, in particular, to higher frequencies, such that a detector such as MAGIS (or AION) with maximum sensitivity in the intermediate-frequency range around $0.1$~Hz would be ideal for mapping out the GW signal and profiling its two major components. For less supercooled transitions and for higher $Z'$ masses the peak moves towards higher frequencies and eventually even the Einstein Telescope could probe the tail of the spectrum.}

Finally, in Fig.~\ref{fig:csigws2} we show the {expected SNR as a function of the model parameters $m_{Z'}$ and $g_{{\rm B}-{\rm L}}$ for the four future observatories. The signal-to-noise ratio is larger than 10 between the black dashed lines.} Above the upper solid black line the bubble wall reaches a terminal velocity before the collisions, and the dominant contribution therefore arises from sound waves in the plasma. Below that line the bubble collisions contribution dominates, and the lower solid line shows where the percolation temperature drops below the critical QCD temperature. The orange line shows where $\Gamma_{\rm dec} = H_*$, and to the right of it the universe experiences a period of matter dominance after the phase transition. {Due to this, the strength of the GW signal rapidly decreases at large $m_Z'$. Above the red line the transition happens before the vacuum energy becomes the dominant energy density component.}

\begin{figure}
\begin{center}
\includegraphics[height=0.38\textwidth]{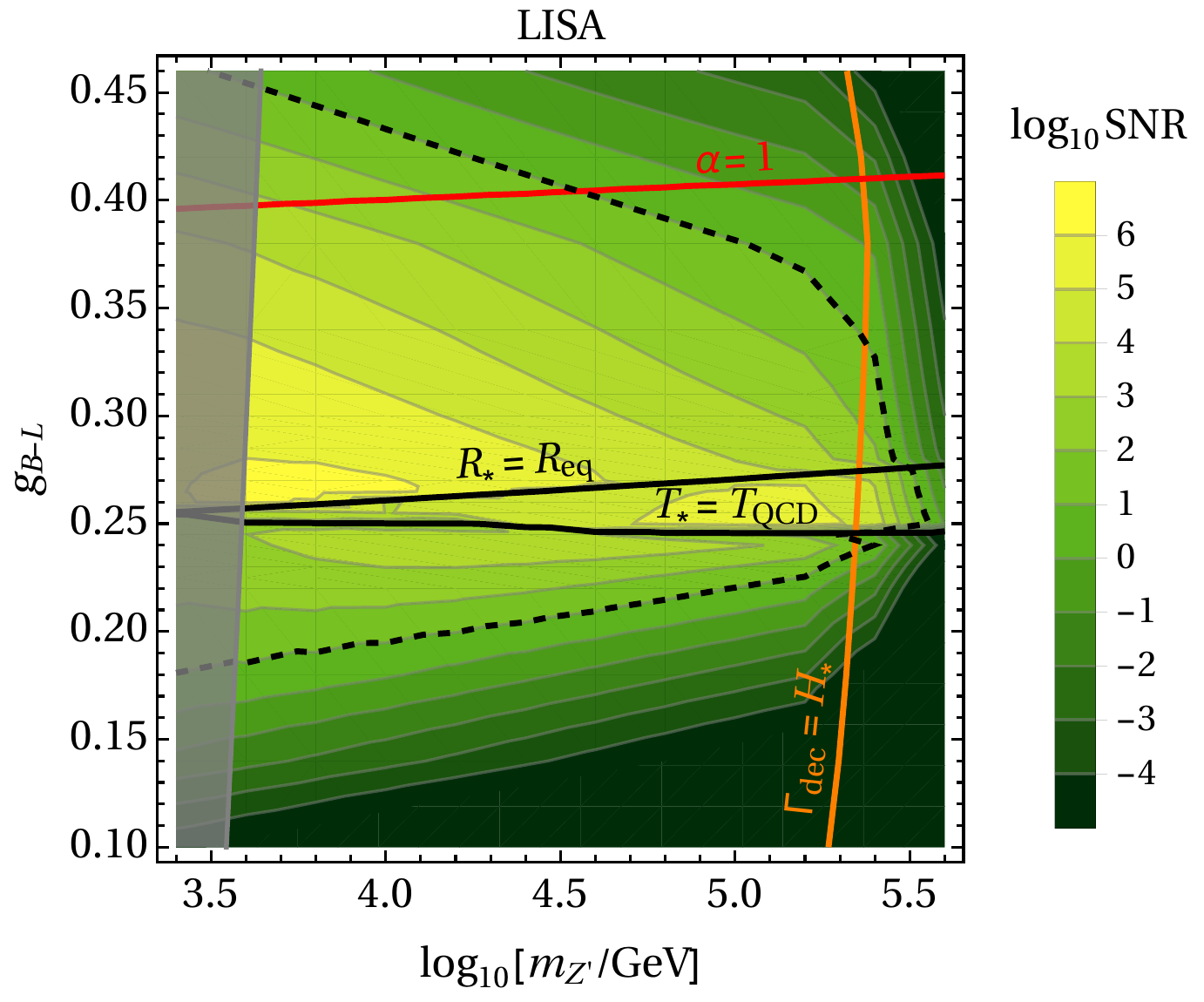} \hspace{2mm}
\includegraphics[height=0.38\textwidth]{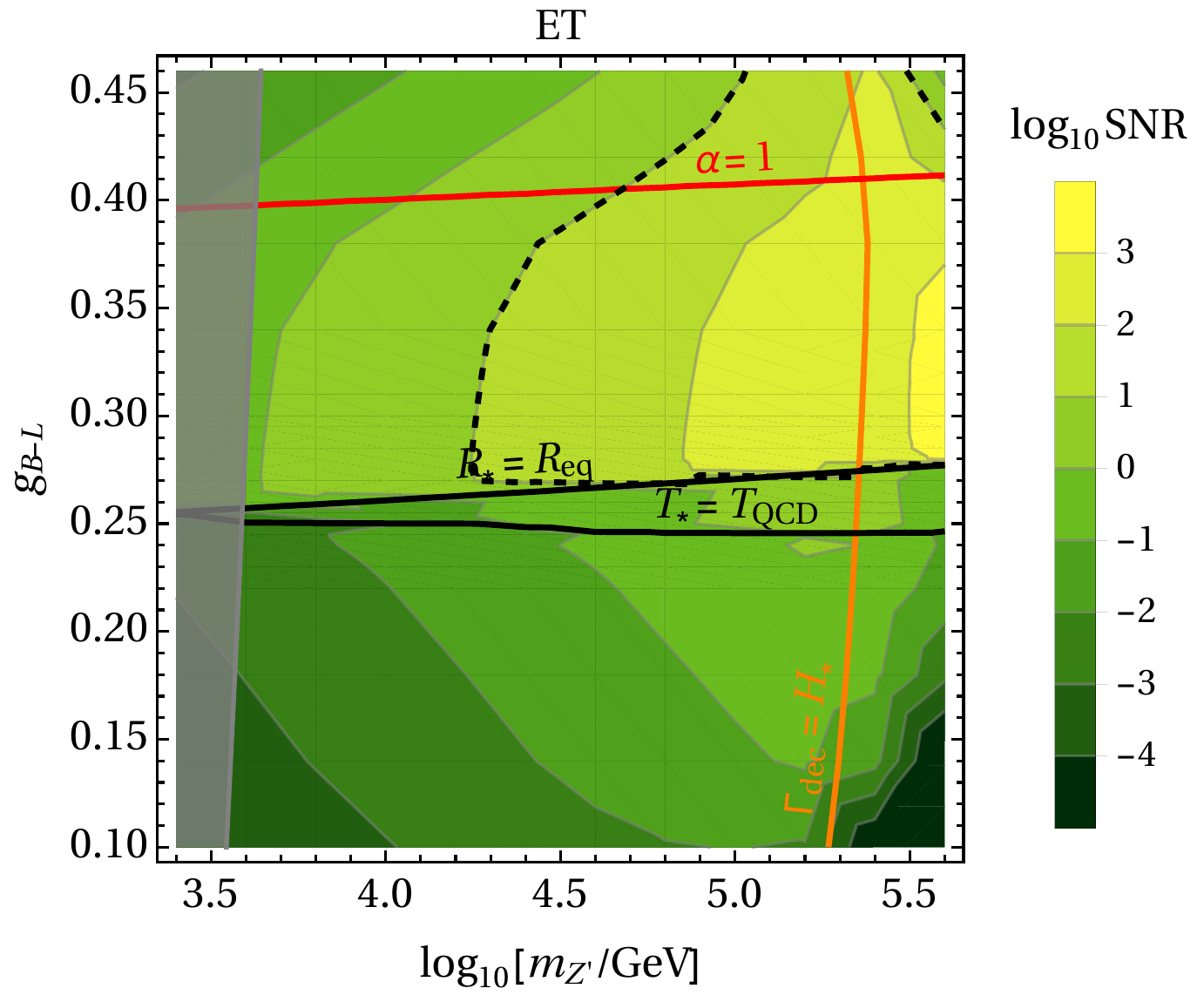} \\ \vspace{2mm}
\includegraphics[height=0.38\textwidth]{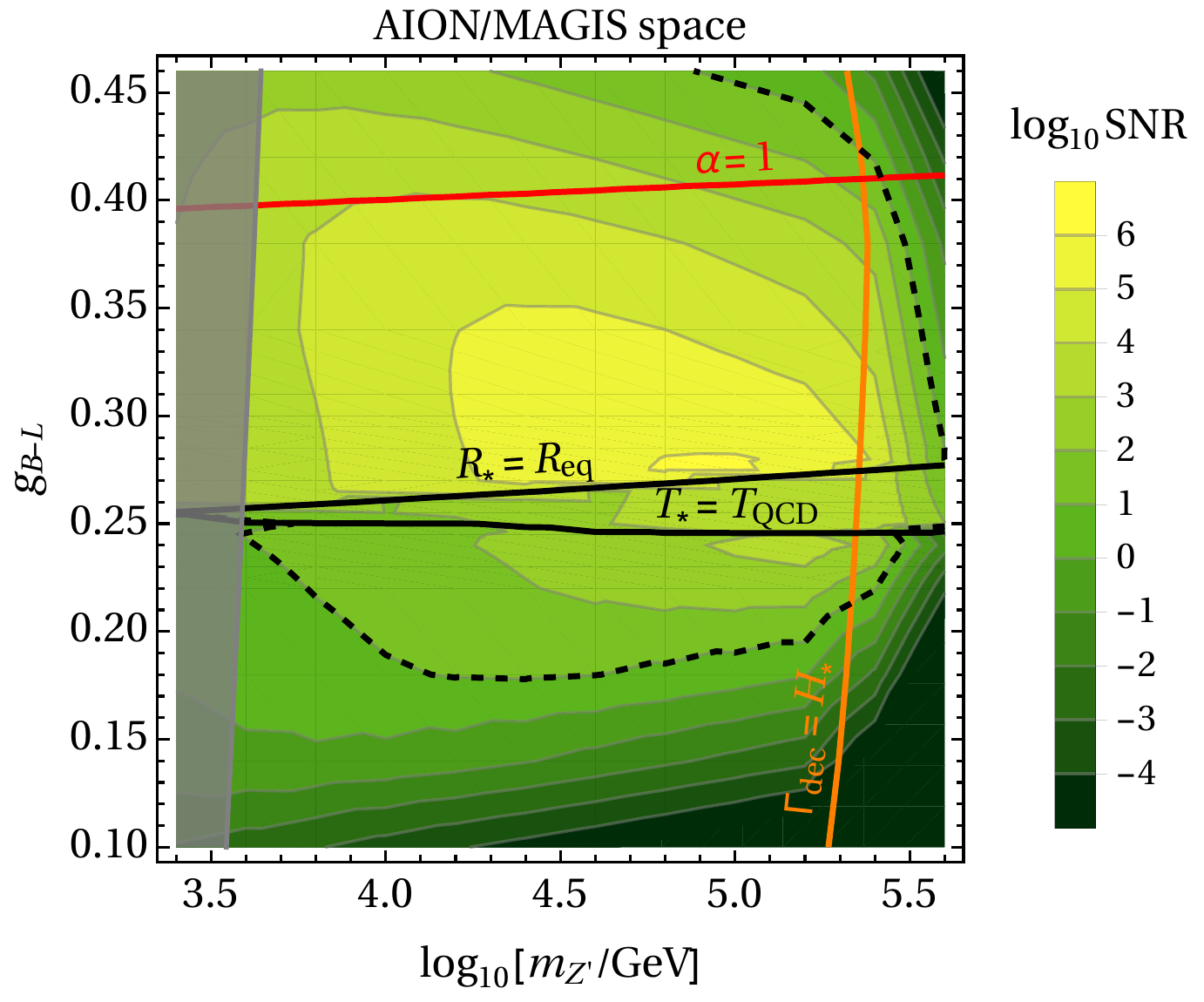} \hspace{2mm}
\includegraphics[height=0.38\textwidth]{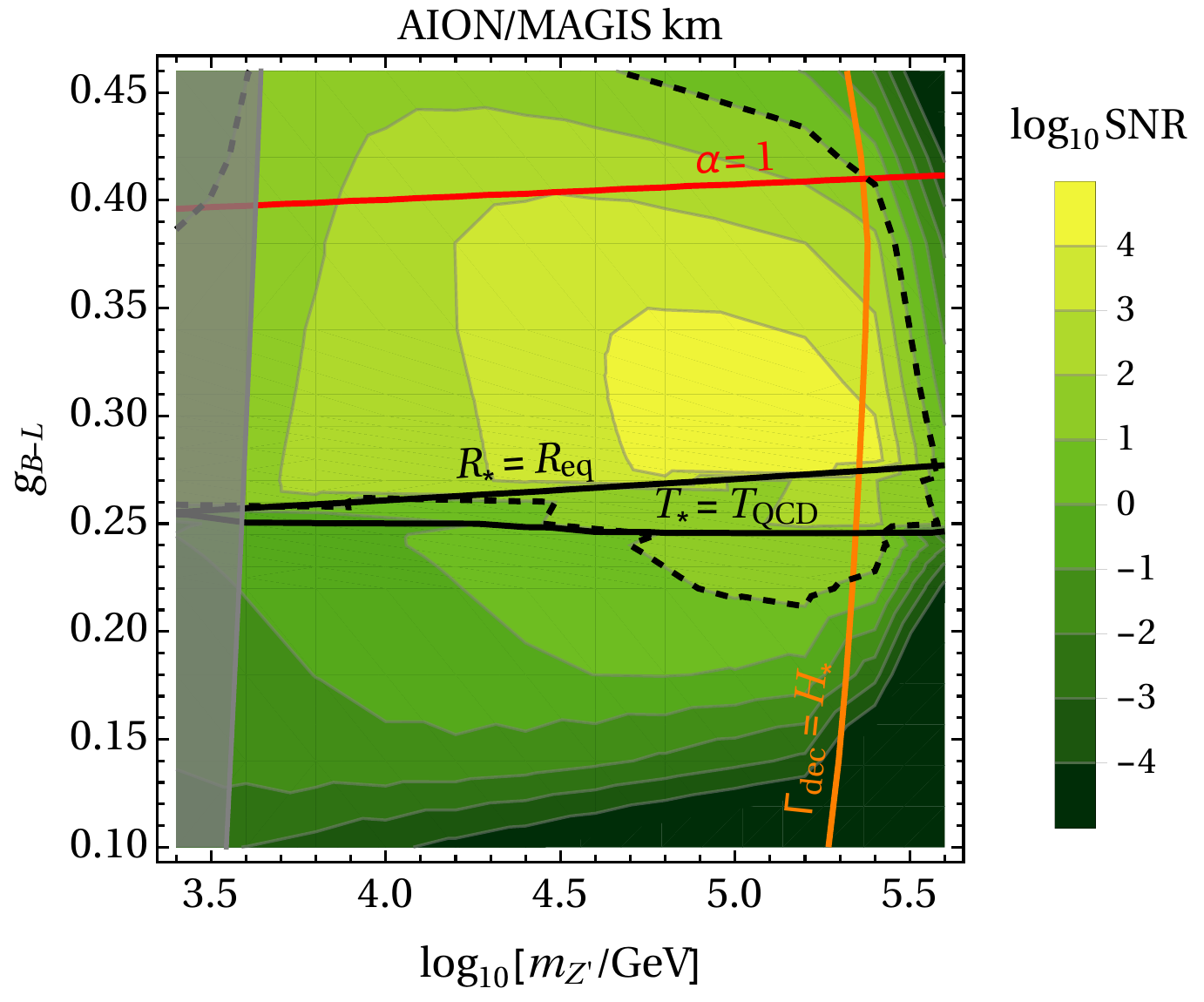}
\caption{The color coding shows the magnitude of the signal-to-noise ratio for {LISA (upper left), ET (upper right), AION/MAGIS space (lower left) and AION/MAGIS km (lower right)} in the classically scale-invariant model. The black dashed line highlights the SNR$=10$ contour. The shaded region on the left is excluded by LHC searches. To the right of the orange line there is a period of matter dominance after the phase transition. Below the upper solid black line the bubble walls reach terminal velocity before collisions and below the lower solid black line the transition happens at $T_*<T_{\rm QCD}$.}
\label{fig:csigws2}
\end{center}
\end{figure}

{Before concluding, we note that in earlier studies of this model~\cite{Jinno:2016knw,Marzo:2018nov} the GW signal was calculated accounting only for the bubble collision signal with efficiency factor $\kappa_{\rm col}=1$, and the signal was compared only to the expected LISA sensitivity. We find that the bubble collisions dominate the signal only at small values of $g_{{\rm B}-{\rm L}}$ (below the $R_*=R_{\rm eq}$ line in Fig.~\ref{fig:csigws2}), in which case the transition is very strongly supercooled. For larger values of $g_{{\rm B}-{\rm L}}$ the signal sourced by plasma motions can be very strong, making in particular that part of the parameter space testable with future GW observatories. Moreover, we show for the first time that this model can be probed also by AION/MAGIS km, AION/MAGIS space and ET.}

\section{Conclusions}
\label{sec:conclusions}

We have calculated in this paper the energy budget of a very strong first-order phase transition. This enabled us to calculate the efficiency factors predicting the amount of energy sourcing the GW spectrum from bubble-wall collisions and also spectra associated with the plasma.
{
We have found that, since shocks {generally} develop in the plasma flow within a Hubble time, the spectra from sound waves and turbulence {are expected to} be of {comparable} order, with sound waves dominating the GW spectrum only in a small region close to its peak. We also have given general formulae to check how much supercooling is needed before the bubble wall contribution to the GW spectra becomes relevant. Using specific examples we have found that only (nearly) classically scale-invariant models allowing for a prolonged inflationary phase before the transition completes can provide enough supercooling to produce bubble-wall-dominated spectra.}

{
We have also performed {numerical} simulations {of expanding bubbles} to elucidate the connection between the initial bubble profiles and their final form upon collision, {and have verified} that in realistic examples most relevant bubbles of size $R_*$ always grow for long enough to evolve into a kink-like configuration leading to predictions resembling those {of} the thin-wall approximation. This simplifies significantly the calculation of the final energy budget of the phase transition.}
{
{In addition}, we have described in detail the equation of state and the resulting cosmological evolution during the transition. We have verified that the more complete treatment of the equation of state confirms simplified predictions involving only the primordial plasma and constant vacuum energy contribution to the Hubble rate around the percolation temperature.}

We have used these general prescriptions to sharpen predictions in two classes of models of particular interest. Specifically, we have discussed the modification of the SM Higgs potential by the inclusion of an $|H|^6$ non-renormalisable operator. Secondly, we turned to models featuring classical scale invariance, which are capable of realizing scenarios with very significant supercooling. 
{In the case of a modification of the SM potential with the non-renormalisable $|H|^6$ operator} we have confirmed and quantified more precisely that the resulting bubble collision spectrum would be {highly} subdominant with respect to plasma-related sources (by at least ten orders of magnitude) in all the relevant parameter space. We have also explored how the more detailed estimate of the equation of state changes the parameter space, finding that the parameter space in which percolation is guaranteed is not changed, as the probability of remaining in the false vacuum is not significantly altered at the percolation temperature.
In the case of models with a classically scale-invariant potential, we have focused on the particular example of the conformal $U(1)_{\rm B-L}$ extension of the SM. We have ascertained for the first time the energy budget of GW sources in the parameter space of the model and specified the amount of supercooling necessary to  produce a sizable bubble collision signal. We have also showed that, just as in the {previous $|H|^6$ scenario,} for all of the parameter space where sound waves are the dominant source of GWs,  nonlinear effects will cut the sound wave period short of a Hubble time and an extra reduction factor $H R_*/U_f < 1$ has to be included in the final results, {leading also to an enhancement of the turbulence GW component}. We also identify the temperature for which the sound wave and collision spectra are comparable, which could lead to an interesting spectrum with two distinct peaks.

Finally, we also note that in many cases where the sound wave spectrum peaks within the LISA band, the part of the spectrum sourced by turbulence has significant extra support outside the region where LISA is most sensitive. Exploring these regions would be important for probing these two components produced by the evolution of the plasma in the linear and non-linear regimes, {and highlights the importance of complementarity between LISA and other GW observatories (see e.g.~\cite{Figueroa:2018xtu}). Incidentally, this strongly} enhances the interest of possible experiments sensitive to frequencies between $10^{-2}$ and 1\,Hz, such as MAGIS~\cite{Graham:2017pmn} and the similar AION proposal~\cite{AION:2018} as well as the DECIGO~\cite{Kawamura:2006up} and BBO~\cite{Yagi:2011wg,Crowder:2005nr} proposals.

\acknowledgments

The work of JE, ML and VV was supported by the UK STFC Grant ST/P000258/1. JE was also supported by the Estonian Research Council via a Mobilitas Pluss grant and ML by the Polish MNiSW grant IP2015 043174. JMN was supported by the Programa Atracci\'on de Talento de la Comunidad de Madrid via grant 2017-T1/TIC-5202 and the Spanish MINECO's ``Centro de Excelencia Severo Ochoa" Programme via grant SEV-2016-0597.

\bibliographystyle{JHEP}
\bibliography{EWPT}
\end{document}